\documentclass[aps,twocolumn,nofootinbib]{revtex4-2}

\usepackage{graphicx}
\usepackage{dcolumn}
\usepackage{bm}
\usepackage{array}
\usepackage{booktabs}
\usepackage{multirow}
\newcommand{\head}[1]{\textnormal{\textbf{#1}}}
\newcommand{\normal}[1]{\multicolumn{1}{l}{#1}}
\usepackage{amssymb,amsfonts}
\usepackage{epsfig}
\usepackage{epstopdf}
\usepackage[all]{xy}
\usepackage{amsthm}
\usepackage{dcolumn}
\usepackage{url}
\usepackage{dsfont}
\usepackage{slashed}
\usepackage{mathrsfs}

\newcommand{\be}{\begin{equation}}
\newcommand{\ee}{\end{equation}}
\newcommand{\bea}{\begin{eqnarray}}

\newcommand{\eea}{\end{eqnarray}}

\def\inbar{\,\vrule height1.5ex width.4pt depth0pt}
\def\IR{\relax{\rm I\kern-.18em R}}
\def\IC{\relax\hbox{$\inbar\kern-.3em{\rm C}$}}

\begin{document}
\title{Heavy quarks structure functions $F^{q\bar{q}}_{2}(x, Q^{2})$ and $F^{q\bar{q}}_{L}(x, Q^{2})$ at the NLO approximation}

\author{S. Zarrin\footnote{zarrin@phys.usb.ac.ir}}

\author{S. Dadfar}

\affiliation{ Department of Physics, University of Sistan and Baluchestan, Zahedan, Iran}

\begin{abstract}
We provide compact formulas for the heavy quarks structure functions  $F^{q\bar{q}}_{2}(x, Q^{2})$ and  $F^{q\bar{q}}_{L}(x, Q^{2})$, with $q=c,b$ and $t$, in $e^{-}p$ interaction  with respect to the behavior of the gluon density at the next-to-leading order (NLO) approximation and present ratios $R^{q\bar{q}}=F^{q\bar{q}}_{L}/ F^{q\bar{q}}_{2}$ for these quarks. To do this, we obtain a fitted form of the heavy quark coefficient functions for deep-inelastic lepton-hadron scattering at a wide range of $Q^{2}$ values. The obtained numerical results  are compared with experimental data from HERA and with the results from the KLN model and the MSTW2008 predictions.
\newline

PACS numbers: 14.65.Dw, 14.65.Fy, 14.65.Ha
 
\end{abstract}
\maketitle
\section{Introduction}

One of the important areas of research at present and future accelerators is the study of the heavy quarks production and it is the most important test of quantum chromodynamics (QCD). These quarks can be produced in the hadron-hadron, photon-hadron, electron-positron and lepton-hadron interactions.  In the QCD framework, the heavy quarks are produced in two different prescriptions. The first one is the so-called variable-flavour number scheme (VFNS) \cite{1}. In this framework, the heavy flavour quarks are treated as a massless quark and their contributions are described by a parton density in the hadron. In the ‘massless’ scheme, at the LO approximation, the dominant contribution is due to the quark parton model (QPM) process. At the NLO approximation, the photon-gluon fusion (PGF) and QCD Compton  processes also contribute. The second one was advocated in Refs. \cite{2,3} where the heavy quarks are treated as massive quark and their contributions are given by fixed-order perturbation theory (FOPT).  In the ‘massive’ scheme, the dominant LO process is the PGF and the NLO diagrams are of order $\alpha_{s}^{2}$.  

At HERA the dominant process for the heavy quarks production in electron-proton interaction $e^{-}+p\rightarrow q\bar{q}+e^{-}+X$ (where $q$ can be $c$, $b$, $t$) (at leading order) is the PGF. In this process, a heavy quark-antiquark pair is formed by the interaction of a virtual photon emitted by the incoming electron with a gluon from the proton. The HERA data show that the production of these quarks is sensitive to the gluon distribution (which the minimum momentum fraction of gluon $x_{g}$ in photoproduction to produce a heavy quark pair is  arranged  such  that $x^{tt}_{g} > x^{bb}_{g }> x^{cc}_ {g}$)  and also depends on the mass  of these quarks. So,  the  calculations of the heavy quarks structure functions depend  on  a  wide  range  of  perturbative  scales $\mu^{2}$ \cite{4,5,6,7,8}. Theoretical calculations at the LO and NLO processes for producing the heavy quarks ($c$ and $b$) are available at HERA.  At small $x$,  both the H1 and ZEUS detectors have measured the charm and beauty components  $F^{c\bar{c}}_{2}$ and  $F^{b\bar{b}}_{2}$ of the structure functions with the $e^{-}p$  center-of-mass  energies $\sqrt{s}=319GeV$ and $\sqrt{s}=318GeV$, respectively  \cite{4,5}.  These structure functions are obtained from the measurement of  the  inclusive  heavy quark cross-sections after applying small corrections to  the  heavy quark  longitudinal  structure  function at low and moderate  inelasticity.
 
In high energy processes, the heavy quark contribution to the proton structure functions will be studied in projects such as the Large Hadron electron Collider (LHeC) and the Future Circular Collider electron-hadron (FCC-eh) which run to beyond a TeV in center-of-mass energy \cite{9,10,11,12,13,14}. At the LHeC project, the possibility of colliding an electron beam from a new accelerator with the existing LHC proton is investigated.  In this project,  the  electron-proton  center  of  mass  energy  is planned to reach $\sqrt{s}=1.3TeV$  \cite{9,13}. The FCC-eh project is an ideal environment to increase center-of-mass energy.  In the proposed FCC-eh programme, the  heavy quark  distributions  will be examined at $\sqrt{s}=3.5TeV$ \cite{14}.

Theoretically, in Refs. \cite{15,16,17,18,19}, the connection between the gluon distribution and the heavy quarks (charm  and beauty) structure  functions have been shown at small $x$. Moreover, in Refs. \cite{20,21} the author presents the conditions necessary to predict the top structure function $F^{t\bar{t}}_{2}$ with respect to the different predictions for the behavior of the gluon at low $x$ and high  $Q^{2}$ values. Furthermore, recently various successful phenomenological methods have presented to obtain the charm and beauty structure functions and the ratios of $R^{c\bar{c}}$ and $R^{b\bar{b}}$ \cite{16,21,22}. The importance of these studies,  along  with  the  $t$-quark density, can  be  explored  at  future circular collider energies  and may lead us to new physics in the future \cite{23,24}. As mentioned, at HERA, the heavy quarks longitudinal structure function is considered as a small correction at low and moderate inelasticity to calculate the heavy quark production cross section. But, in the region of high inelasticity, this function may has a significant effect on the heavy quark production cross section.  The heavy quark reduced cross section is written in terms of the heavy quark structure functions as:

\begin{equation}\label{eq:1}
\tilde{\sigma}^{q\bar{q}}(x,Q^{2})=F_{2}^{q\bar{q}}(x,Q^{2})\bigg[1-\frac{y^{2}}{Y_{+}}\frac{F_{L}^{q\bar{q}}(x,Q^{2})}{F_{2}^{q\bar{q}}(x,Q^{2})}\bigg],
\end{equation}
where  $y=Q^{2}/(xs)$ is the inelasticity variable in which $s$  and $Q^{2}$ are the center-of-mass energy squared and the photon virtuality, respectively, and  $Y_{+}=1+(1-y)^{2}$. The heavy quark structure functions $F_{2}^{q\bar{q}}(x,Q^{2})$ and $F_{L}^{q\bar{q}}(x,Q^{2})$ with respect to the behavior of the gluon density are given by:
\begin{equation}
F_{k,g}^{q\bar{q}}(x,Q^{2},m_{q}^{2})=xe_{H}^{2}\int_{x}^{z_{max}}H_{k,g}(z,\xi) g(\frac{x}{z},\mu^{2})\frac{dz}{z},
\end{equation}
where $\mu^2=(Q^2+4m_{q}^{2})^{1/2}$ is the default common value for the factorization and renormalization scales, $z_{max}=\frac{Q^{2}}{4m_{q}^{2}+Q^{2}}$ and $\xi=\frac{Q^{2}}{m_{q}^2}$. In general, the heavy quark coefficient functions of $H_{k,g}(z,\xi)$ (with $k=2,L$) are expanded in $\alpha_{s}$ as follows:
\begin{equation}
H_{k,g}(z,\xi)=\sum_{i=1}^{\infty}\left(\frac{\alpha_{s}(\mu^{2})}{4\pi}\right)^{i}H^{(i)}_{k,g}(z,\xi), \ \ k=2,L,
\end{equation}
where the heavy quark coefficient functions at the LO and NLO approximations, $H_{k,g}^{(1)}$ and $H_{k,g}^{(2)}$, are as follows:
$$
H_{k,g}^{(1)}(z,\xi)=\frac{\xi}{\pi z}c^{(0)}_{k,g}(\eta,\xi),
$$
\begin{equation}
H_{k,g}^{(2)}(z,\xi)=\frac{16 \pi \xi}{ z}\left[c^{(1)}_{k,g}(\eta,\xi)+\bar{c}^{(1)}_{k,g}(\eta,\xi)\ln\left(\frac{\mu^{2}}{m_{q}^{2}}\right)\right],
\end{equation}
where the coefficient functions $c^{(0)}_{k,g}(\eta,\xi)$ have been given in Ref. \cite{3}. The coefficients $c^{(1)}_{k,g}$ and $\bar{c}^{(1)}_{k,g}$ are rather lengthy, and not published in print and they are only available as computer codes \cite{3}. In Ref. \cite{25}, the analytic form of the heavy quark coefficient functions have been presented  for deep-inelastic lepton-hadron scattering in the kinematical regime $Q^{2}\gg m^{2}_{q}$ in which $Q^{2}$ and $m^{2}_{q}$ stand for the masses squared of the virtual photon and heavy quark, respectively. 

In  this paper, in section II, we modify  the presented heavy quark coefficient functions in Ref. \cite{25} (where have been obtained by using the two-loop operator matrix elements up to non-logarithmic terms and the NLO light parton coefficient functions) with a suitable fit by using the heavy quark structure functions from HERA \cite{26,27,28,281},  LHeC \cite{292} and other works as Ref. \cite{291} (for $c$ and $b$ quarks) and Refs. \cite{20,21} (for $t$ quark).  We show that these coefficient functions are usable in a wide range of photon virtuality of $Q^{2}$ values. Then, by using DGLAP evolution equations and the Laplace transform method,  we obtain the gluon distribution function and consequently, by using Eq. (2), present the heavy quark structure functions $F_{2}^{q\bar{q}}$ and  $F_{L}^{q\bar{q}}$ at the LO and NLO approximations. In section III, we compare our numerical results with HREA data \cite{26,27,28,281}, with the results from the KLN model \cite{29} and Ref. \cite{22} and the MSTW2008 predictions \cite{32}. 
\section{METHOD}

Let us first present the heavy quark coefficient functions in low $Q^{2}$. The presented heavy quark coefficient functions (at the NLO approximation) in Ref. \cite{25} at the regimes of $Q^2\geq m^{2}$ and  $Q^2\leq m^{2}$ do not provide appropriate and acceptable results and the obtained structure functions by these coefficient functions are very ascendant, especially for the charm quark. To control this growth, by using HERA data \cite{26,27,28,281} and the results from Refs. \cite{20,21,291,292}, we obtain a series of control coefficients that are only a function of $Q^{2}$. By multiplying these coefficients by the obtained coefficient functions in Ref. \cite{25}, they give acceptable results. The general form of these control coefficients is as $b-\exp(cQ^{2})$, which $b$  and $c$ are  fixed numbers and are shown in table (1) for the heavy quarks. In Figure (1), we compare the fitted functions of  $H_{2,g}^{(2)}$, $H_{L,g}^{(2)}$  for the charm quark with Ref. \cite{25}. As can be seen, by applying these control coefficients, the heavy quarks coefficient functions reduce at low $Q^{2}$. 
 
Now, we intend to obtain the gluon distribution function in terms of $Q^{2}$ and $x$. For this aim, we use the DGLAP evolution equations and the Laplace transform method as Ref. \cite{31}. The coupled DGLAP integral-differential equations are as follows \cite{30}:
$$
 \frac{\partial F_{s}(x,Q^{2})}{\partial \ln Q^{2}}=\frac{\alpha_{s}(Q^{2})}{2\pi}\bigg[P_{qq}(x,Q^{2})\otimes F_{s}(x,Q^{2})
 $$
\begin{equation}
 +2n_{f}P_{qg}(x,Q^{2})\otimes G(x,Q^{2})\bigg],\
\end{equation}
$$
\frac{\partial G(x,Q^{2})}{\partial \ln Q^{2}}=\frac{\alpha_{s}(Q^{2})}{2\pi}\bigg[P_{gq}(x,Q^{2})\otimes F_{s}(x,Q^{2})
$$
\begin{equation}
+P_{gg}(x,Q^{2})\otimes G(x,Q^{2})\bigg], \quad
\end{equation}
where $\alpha_{s}$ is the running strong coupling constant and $P_{ab}(x)$'s  are the Altarelli-Parisi splitting functions which have the following form:
\begin{equation}
P_{ab}(x,Q^{2})=P^{(0)}_{ab}(x)+\frac{\alpha_{s}(Q^{2})}{2\pi}P^{(1)}_{ab}(x)+...  \ .
\end{equation}
The standard representation of the QCD running coupling constant in the LO and NLO (within the MS-scheme) approximations have the forms:
\begin{equation}
\alpha_{s}^{LO}(t)=\frac{4\pi}{\beta_{0}t},\qquad\qquad\qquad
\end{equation}
\begin{equation}
\alpha_{s}^{NLO}(t)=\frac{4\pi}{\beta_{0}t}\left(1-\frac{\beta_{1}\ln t}{\beta_{0}^{2}t}\right),
\end{equation}
 \begin{table}[h]
\begingroup
\fontsize{10pt}{12pt}\selectfont{
\label*{ Table (1): The fixed numbers $b$ and $c$  in the control coefficients of the heavy quark coefficient functions. }}
\newline
\endgroup
\begingroup
\fontsize{8pt}{12pt}\selectfont{
\begin{tabular} {ccc}
\toprule[1pt]
\quad\quad\quad\quad&\normal{\head{$b $}}&
\normal{\head{$c $}}\  \\ \hline
$m_{c}$\quad \quad & $1.007324$\quad\quad&  $-11.9235\times 10^{-4}$   \\ \\
$m_{b}$ \quad\quad& $1.0036687$\quad\quad &  $-2.35842\times 10^{-4}$   \\ \\  
$m_{t}$\quad\quad & $1.000011$ \quad\quad&  $-8.93697\times 10^{-8}$   \\ \\  
    
\bottomrule[1pt]
\end{tabular} }
\endgroup
\end{table}
\begin{figure}[h]
\begin{center}
\includegraphics[width=.5\textwidth]{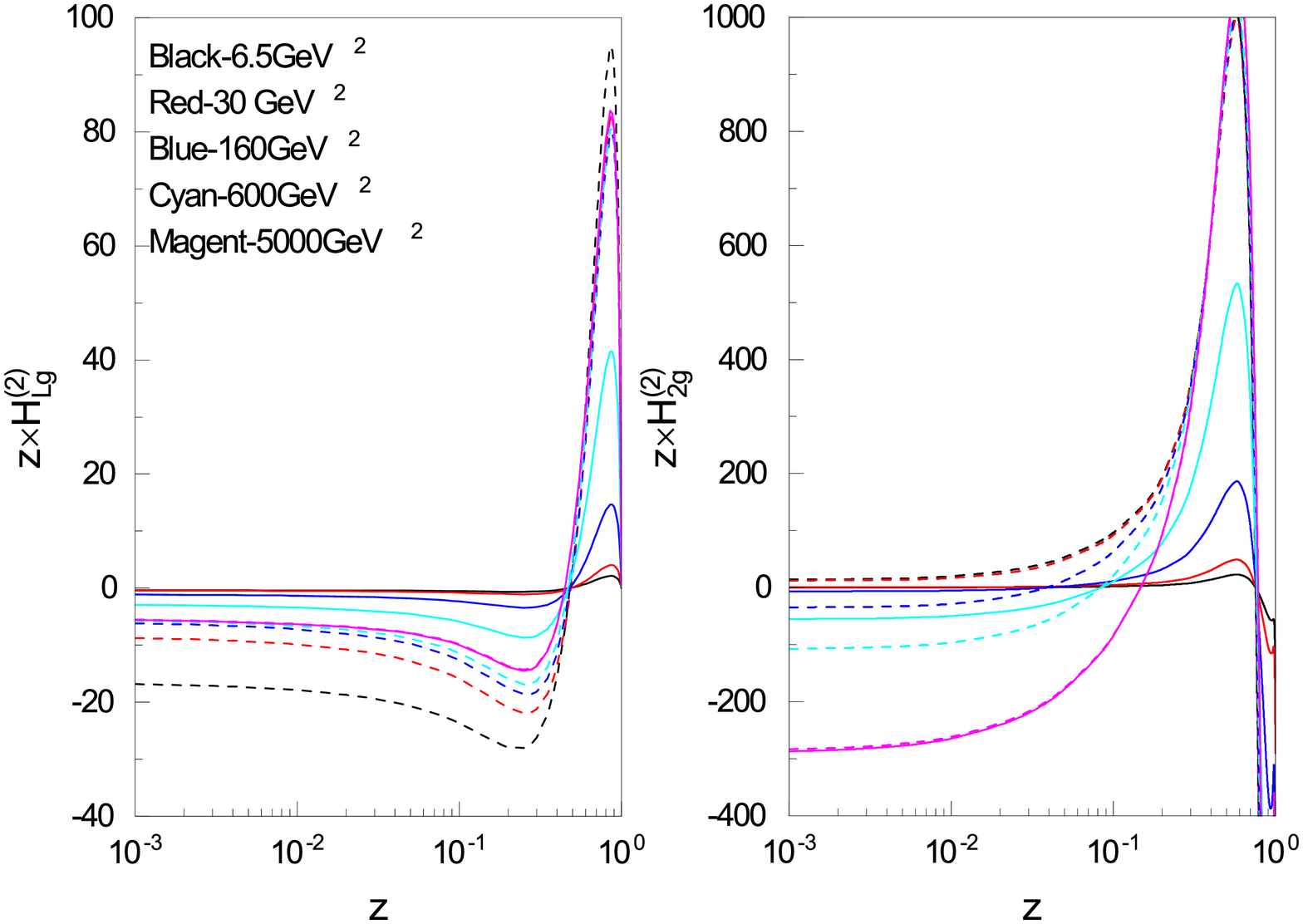}
\end{center}
\begin{center}
\selectfont{\label*{Figure (1): A comparison  between the fitted charm quark coefficient function and those presented in Ref. \cite{25}. The solid curves are the fitted results and the dashed curves are correspond to the results of Ref. \cite{25}.}}
\end{center}
\end{figure}
where $\beta_{0}=\left(11-2/3n_{f}\right)$, $\beta_{1}=\left(102-38/3n_{f}\right)$ and $t=\ln(Q^{2}/\Lambda^{2})$ in which $\Lambda$ is
the QCD cut-off parameter. The symbol $\otimes$ represents the convolution integral which is defined as $f(x)\otimes h(x)=\int_{x}^{1}f(y)h(x/y)dy/y$. To solve Eqs. (5) and (6), we use here the Laplace transform method. For this aim, we insert the variables $x=\exp(-v)$, $y=\exp(-w)$ and $\tau(Q^{2},Q_{0}^{2})=\frac{1}{4\pi}\int_{Q^{2}_{0}}^{ Q^{2}} \alpha_{s}(Q{'}^{2})d\ln(Q{'}^{2})$ into the DGLAP evolution equations as follows:
$$
 \frac{\partial \hat{F}_{s}(v,\tau)}{\partial\tau}=2\bigg[\int_{0}^{v}\hat{P}_{qq}(v-w,\tau) \hat{F}_{s}(w,\tau)dw
$$
 \begin{equation}
 +\int_{0}^{v}4n_{f}\hat{P}_{qg}(v-w,\tau)\hat{G}(w,\tau)dw\bigg],\
\end{equation}
$$
\frac{\partial \hat{G}(v,\tau)}{\partial\tau}=2\bigg[\int_{0}^{v}\bigg[\hat{P}_{gq}(v-w,\tau) \hat{F}_{s}(w,\tau)dw
$$
\begin{equation}
+\int_{0}^{v}\hat{P}_{gg}(v-w,\tau) \hat{G}(w,\tau)dw\bigg], \quad
\end{equation}
In above equations $\hat{P_{ij}}(v,\tau)\equiv P_{ij}(\exp(-v),\tau)$, $\hat{F_{s}}(v,\tau)\equiv F_{s}(\exp(-v),\tau)$ and $\hat{G}(v,\tau)\equiv G(\exp(-v),\tau)$
The convolution theorem for the Laplace transforms allows us to rewrite the right-hand sides of Eqs. (10) and (11) with considering the fact that the Laplace transform of the convolution factors is simply the ordinary product of the Laplace transform of the factors. Using the Laplace transform method, we can turn the convolution equations at the LO and NLO approximations from $v$-space into $s$-space, and then solve them straightforwardly in $s$-space as:
\begin{equation}\label{eq:11}
f^{(i)}(s,Q^{2})=k_{ff}^{(i)}(s,\tau)f^{(i)}(s,Q_{0}^{2})+k_{fg}^{(i)}(s,\tau)g^{(i)}(s,Q_{0}^{2}),
\end{equation}
\begin{equation}\label{eq:12}
g^{(i)}(s,Q^{2})=k_{gf}^{(i)}(s,\tau)f^{(i)}(s,Q_{0}^{2})+k_{gg}^{(i)}(s,\tau)g^{(i)}(s,Q_{0}^{2}),
\end{equation}
with $i=$ LO or NLO. The functions of $f(s,0)$ and $g(s,0)$ are the singlet and gluon distribution functions at initial scale $\tau=0$ (i.e., $Q^{2}_{0}$). In Eqs. (\ref{eq:11}) and (\ref{eq:12}), the kernels of $k_{ij}(s,u)$'s at the LO and NLO approximations can be found in Ref. \cite{31} and $\mathcal{L}[\hat{H}(v,\tau),v,s]= h(s,\tau)$. Since the gluon distribution functions in the above equation is in Laplace space $s$ and its exact solution is not possible through analytical techniques, so its inverse Laplace transform must be computed numerically \cite{301,302}.  To obtain the heavy quarks structure functions in terms of the distribution functions at the initial scale, we have to turn Eq. (2) to the Laplace space $s$. To do this, we use variable variable $z=x/y$ and the transformation $x\rightarrow xe^{(\ln 1/a)}$ (where $a$ is larger than one), so we have:

$$
F_{k,g}^{q\bar{q}}(xe^{(\ln 1/ a)},Q^{2},m_{q}^{2})=e_{H}^{2}\int_{x}^{1}G(y,\mu^{2})\frac{dy}{y}
$$
\begin{equation}
\times C_{k,g}(\frac{xe^{(\ln 1/ a)}}{y},\xi), \quad k=2, L,
\end{equation}
where  $G(y,Q^{2})=yg(y,Q^{2})$ and $C_{k,g}(x,\xi)=xH_{k,g}(x,\xi)$. By using the variables $x=\exp(-v)$, $y=\exp(-w)$, one can rewrite the above equation as: 

$$
\hat{F}_{k,g}^{q\bar{q}}(v-\ln 1/a,Q^{2},m_{q}^{2})=e_{H}^{2}\int_{0}^{v}\hat{G}(w,\mu^{2})
$$
\begin{equation}
\times \hat{C}_{k,g}(v-w-\ln 1/a,\xi)dw, \quad k=2, L,
\end{equation}
Using the Laplace transform method, we can turn the above equation from $v$-space into $s$-space as follows:
\begin{equation}
f_{k,g}^{q\bar{q}}(s,Q^{2},m_{q}^{2})=e_{H}^{2}g(s,\mu^{2})h_{k,g}(s,\xi), \quad k=2,L, 
\end{equation}
where $ h_{k,g}(s,\xi)=\mathcal{L}[\hat{C}_{k,g}(v-\ln 1/a,\xi)],v,s]$. It should be noted that the coefficients $H^{(1)}_{k,g}(x,\xi)$ cannot be transferred to the Laplace space with their exact form, in order to be able to easily transfer them to the Laplace space, we expand them in a suitable way and present them in the Appendix. The results of these expansions are numerically shown and compared with exact values of the coefficients $H^{(1)}_{k,g}(\eta,\xi)$ until $z_{max}$ in figure (2). At the bottom of this figure, we show the ratio of the approximate results to exact results. The coefficients $h_{k,g}(s,\xi)$ at the LO approximation are presented in Appendix but at the NLO approximation become too lengthy to reproduce here, it is easily calculated (like the LO approximation) using a program such as Mathematica or Maple. To obtain the heavy quarks components $F^{q\bar{q}}_{2}$ and $F^{q\bar{q}}_{L}$ of the structure functions in Laplace space at the LO and NLO approximations, we insert the obtained gluon distribution function in Eq. (13) into Eq. (16). But before that, $Q^{2}$ must be replaced by $\mu^{2}$.  With these descriptions, we can write the structure functions as follows:
$$
f_{k,g}^{q\bar{q}}(s,Q^{2},m_{q}^{2})=e_{H}^{2}\Bigg[j_{gf}(s,\mu^{2})f(s,Q_{0}^{2})
$$
\begin{equation}
+ j_{gg}(s,\mu^{2})g(s,Q_{0}^{2})\Bigg], \quad k=2, L,
\end{equation}
where
$$
j_{gf}(s,\mu^{2})= h_{k,g}(s,\xi)k_{fg}(s,\tau(\mu^{2},Q_{0}^{2}))/a^{s},
$$
\begin{equation}
j_{gg}(s,\mu^{2})= h_{k,g}(s,\xi)k_{gg}(s,\tau(\mu^{2},Q_{0}^{2}))/a^{s}.
\end{equation}
\begin{figure}[h]
\begin{center}
\includegraphics[width=.48\textwidth]{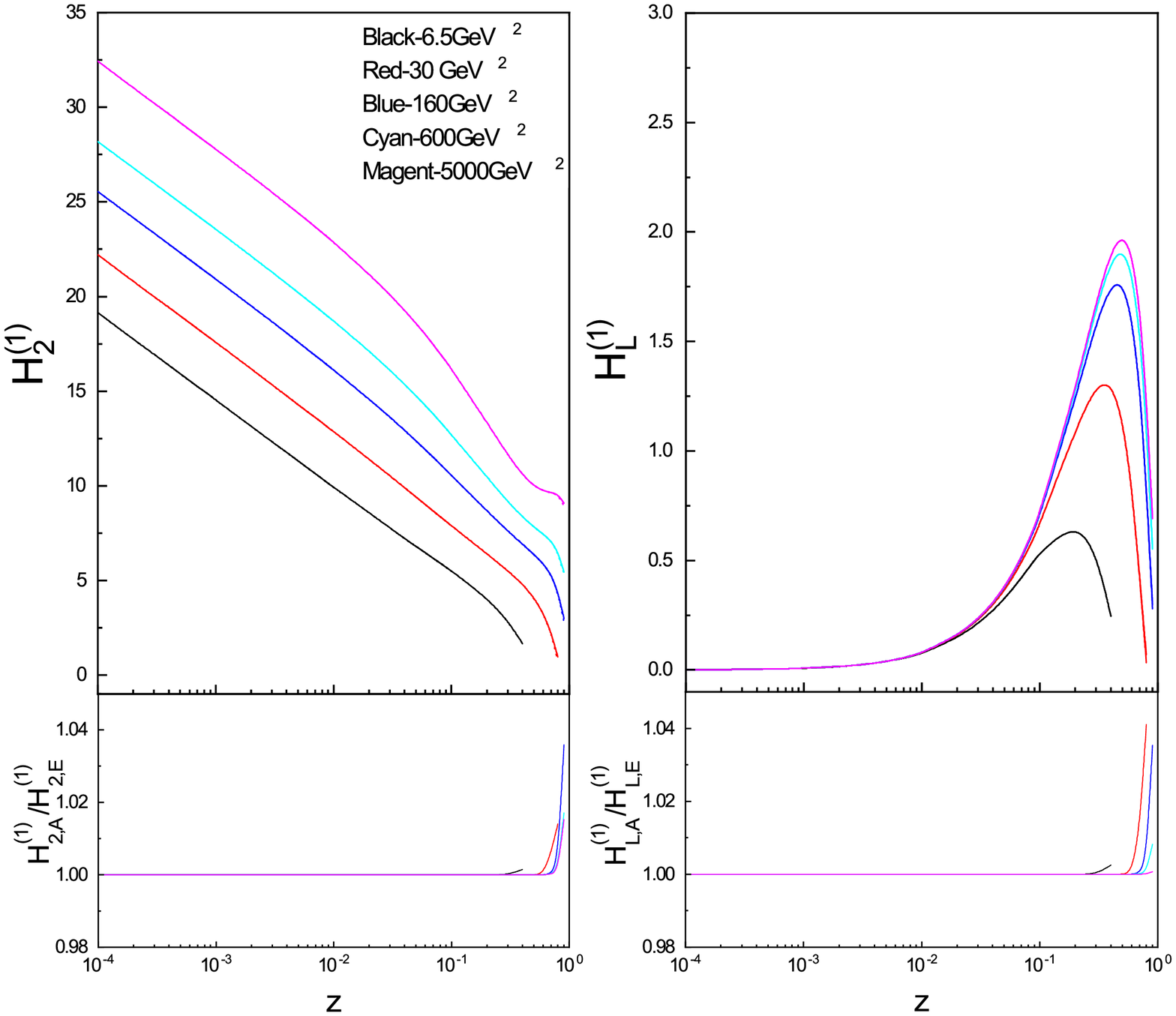}
\end{center}
\begin{center}
\selectfont{\label*{Figure (2): A comparison between exact and approximate values of the coefficient function $H^{(1)}_{k,g}$ for the charm quark. The solid curves are the approximate results and the dotted curves are the exact results.}}
\end{center}
\end{figure}
Finally, using the Laplace inverse transform, we can obtain these structure functions in the usual space $x$ as follows;
$$
F_{k,g}^{q\bar{q}}(x,Q^{2},m_{q}^{2})=e_{H}^{2}\Bigg[J_{gf}(x,\mu^{2})\otimes F_{s}(x,Q_{0}^{2})
$$
\begin{equation}
+ J_{gg}(x,\mu^{2})\otimes G(x,Q_{0}^{2})\Bigg], \quad k=2, L
\end{equation}
where $J_{gf}(x,\mu^{2})=\mathcal{L}^{-1}[j_{gf}(s,\mu^{2}),s,v]\vert_{v=\ln(1/x)}$ and $J_{gg}(x,\mu^{2})=\mathcal{L}^{-1}[j_{gg}(s,\mu^{2}),s,v]\vert_{v=\ln(1/x)}$. It should be noted that, to obtain the heavy quarks structure functions in the above equation, we require only a knowledge of the singlet $F_{s}(x)$ and gluon $G(x)$  distribution functions at the starting value $Q^{2}_{0}$.
\section{NUMERICAL RESULTS}
Now, we present our numerical results of the heavy quark structure functions $F^{q\bar{q}}_{2}(x,Q^{2})$ and $F^{q\bar{q}}_{L}(x,Q^{2})$ at the LO and NLO approximations obtained by the DGLAP evolution equations and Eq. (2).  In order to present more detailed discussions on our findings, the numerical results for the heavy quarks structure functions are compared with HERA data \cite{26,27,28,281}, with the MSTW2008 predictions \cite{32}, the results from the KLN model \cite{29} and Ref. \cite{21,22}. To extract numerical results, we use the published MSTW2008 \cite{32} initial starting functions $F_{s}(x)$ and $G(x)$  at $Q^{2}_{0}= 1 GeV^{2}$. It should be noted that in calculations, the uncertainties are due to the running charm, beauty and top quark masses $m_{c} = 1.29^{+0.077}_{-0.053}$ $GeV$, $m_{b} =4.049^{+0.138}_{-0.118}$ $GeV$  \cite{6} and $m_{t} =173.5^{+3.9}_{-3.8}$ $GeV$ \cite{33}  where the uncertainties are obtained through adding the experimental fit, model and parameterization uncertainties in quadrature.

In figure (3),  we show the numerical results of the charm structure functions $F^{c\bar{c}}_{2}(x,Q^{2})$ and $F^{c\bar{c}}_{L}(x,Q^{2})$ at the LO and NLO approximations at $Q^{2}=6.5, 12, 30, 80, 160$ and $600$ $GeV^{2}$. In this figure, since nowhere is presented data on the heavy quark longitudinal structure function, we only compare $F^{c\bar{c}}_{2}(x,Q^{2})$  with those presented by ZEUS collider \cite{27}. As can be seen, our numerical results at the NLO approximation are closer to the experimental data than the results of the LO approximation. Figure (4) shows the charm structure function $F^{c\bar{c}}_{2}(x,Q^{2})$ compared to data from H1 \cite{26,28}, ZEUS \cite{281} and the MSTW2008 predictions at the NLO approximation \cite{32} at $Q^{2}=11, 60, 130$ and $500$ $GeV^{2}$. 
Figure (5) shows the comparison of our numerical results of the charm structure function $F^{c\bar{c}}_{2}(x,Q^{2})$ with ZEUS data \cite{26} and with the results of KLN model \cite{29} which obtained for the different renormalization scales ($\mu^{2}=4m_{c}^{2}$ and $\mu^{2}=Q^{2}+4m_{c}^{2}$ with $m_{c}=1.2GeV$). 

Figure (6) indicates our numerical results of the beauty structure functions $F^{b\bar{b}}_{2}(x,Q^{2})$ and $F^{b\bar{b}}_{L}(x,Q^{2})$ at the LO and NLO approximations in $Q^{2}=6.5, 12,30,80, 160$ and $600$ $GeV^{2}$. In this figure, we only compare $F^{b\bar{b}}_{2}(x,Q^{2})$  with ZEUS data \cite{27}. In figure (7), we show the predictions for the beauty structure function compared to
the published H1 data \cite{28} and the MSTW2008 predictions \cite{32}. Clearly at the NLO approximation the comparison is good.  Moreover,  in figure (8), we show $F^{b\bar{b}}_{2}(x,Q^{2})$ at $Q^{2}= 12, 60, 200$ and $650$ $GeV^{2}$  in comparison with H1 data \cite{28}  and with the results from KLN model \cite{29}. 

In figure (9), we present the numerical results of the top structure functions $F^{t\bar{t}}_{2}(x,Q^{2})$ and $F^{t\bar{t}}_{L}(x,Q^{2})$ at the LO and NLO approximation at $Q^{2}=6.5, 12,30,80, 160$ and $600$ $GeV^{2}$. As can be seen in this figure, the top longitudinal structure function at the LO approximation is very small relative to the structure function $F^{t\bar{t}}_{2}(x,Q^{2})$. But with increasing energy, this function increases and becomes noticeable. At the NLO approximation in the energy interval $6.5<Q^{2}<600$ $GeV^{2}$, this structure function is zero. The results $F^{t\bar{t}}_{2}(x,Q^{2})$  of our analysis at $Q^{2}\geq m^{2}_{t}$ are shown in figure (10) and compared with the results of Ref. \cite{21} at the NLO approximation. 

In figure (11) is presented a comparison between the heavy quarks structure functions at $Q^{2}=1000,5000$ and $10000$ $GeV^{2}$. In the figure on the left, their $F_{2}^{q\bar{q}}$ structure functions are compared together, at maximum $Q^{2}$ and minimum $x$, the ratios of  $F^{b\bar{b}}_{2}/F^{c\bar{c}}_{2}$ and $F^{t\bar{t}}_{2}/F^{c\bar{c}}_{2}$ are approximately $0.23$ and $0.01$, respectively. In the figure on the right, their $F_{L}^{q\bar{q}}$ structure functions  are compared, at maximum $Q^{2}$ and minimum $x$, the ratios of  $F^{b\bar{b}}_{L}/F^{c\bar{c}}_{L}$ and $F^{t\bar{t}}_{L}/F^{c\bar{c}}_{2}$ are about $0.24$ and $0.0012$, respectively.

In order to assess the significance of and the theoretical uncertainty at the NLO approximation  to the heavy quark structure functions, we indicate in figure (12) the $Q^{2}$ dependence of $R^{c\bar{c}}$, $R^{b\bar{b}}$, and $R^{t\bar{t}}$ evaluated at the NLO approximation from Eq. (19) in comparison with the results from Ref. \cite{22}  (obtained in the high-energy regime, $Q^{2}\gg m^{2}$, at the LO and NLO approximations) and with Ref. \cite{21}.  The very small difference in low energies seen in this comparison is due to the fact that the amount and range of quarks mass in the two articles are different and also due to the used approximation by Ref. \cite{22}. All of the figures clearly show that the extraction procedure provides correct behaviors of the extracted heavy quark structure functions at the LO and NLO approximations. Moreover, it should be noted that although the NLO corrections are very small for values of $x$ where data is available,  but at low $x$ region  these corrections have many effects on the results of the heavy quarks structure functions. Furthermore,  they often allow one to reduce the uncertainties of the predicted results, as one can see by comparing the bands in almost all of the plots presented in the figures.

\begin{figure}[h]
\begin{center}
\includegraphics[width=.7\textwidth]{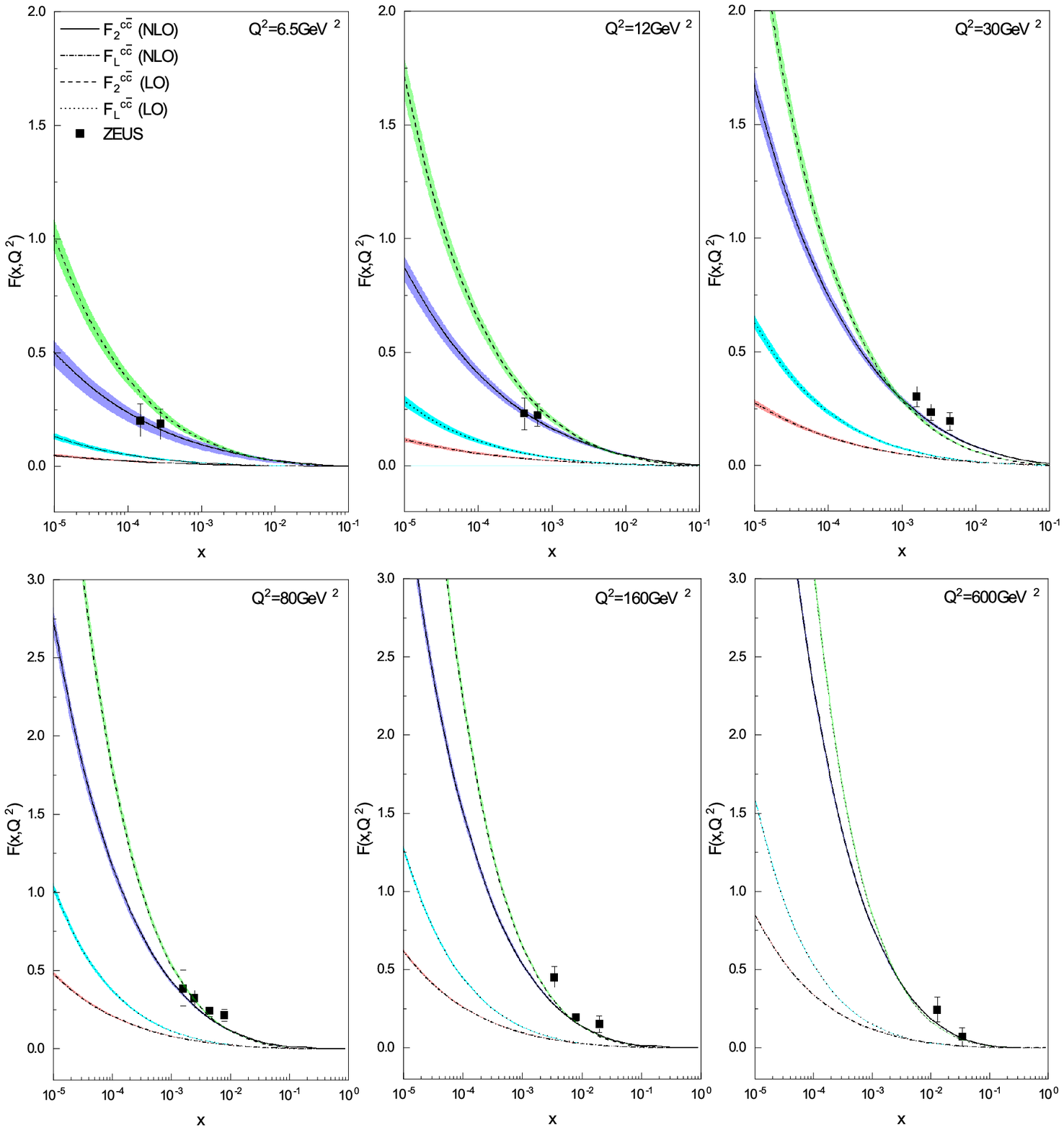}
\end{center}
\begin{center}
\selectfont{\label*{Figure (3): The results of  $F^{c\bar{c}}_{2} $ and $F^{c\bar{c}}_{L} $ at $Q^{2}$ values between $6.5$ and $600$ $GeV^{2}$ as a function of  $x$ at the LO and NLO approximations. The results of $F^{c\bar{c}}_{2} $ are compared with the ZEUS data \cite{27} (solid points). The shaded area are the uncertainties due to the running charm mass.}}
\end{center}
\end{figure}
\begin{figure}[h]
\begin{center}
\includegraphics[width=.48\textwidth]{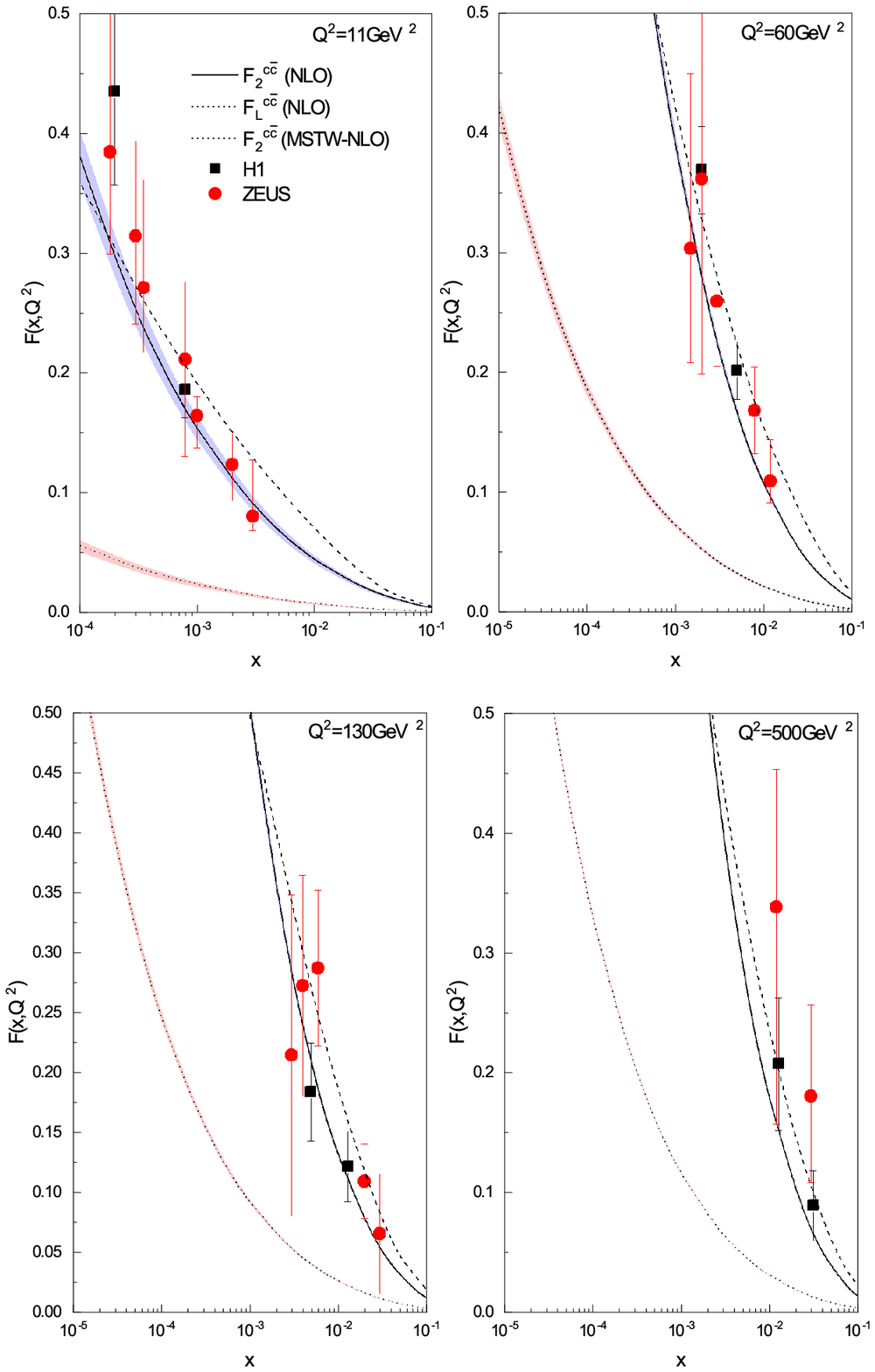}
\end{center}
\begin{center}
\selectfont{\label*{Figure (4): The charm structure function $F^{c\bar{c}}_{2}$ compared to data from H1 \cite{26,28}, ZEUS \cite{281} and MSTW2008 at the NLO approximation \cite{32}. }}
\end{center}
\end{figure}

\begin{figure}[h]
\begin{center}
\includegraphics[width=.5\textwidth]{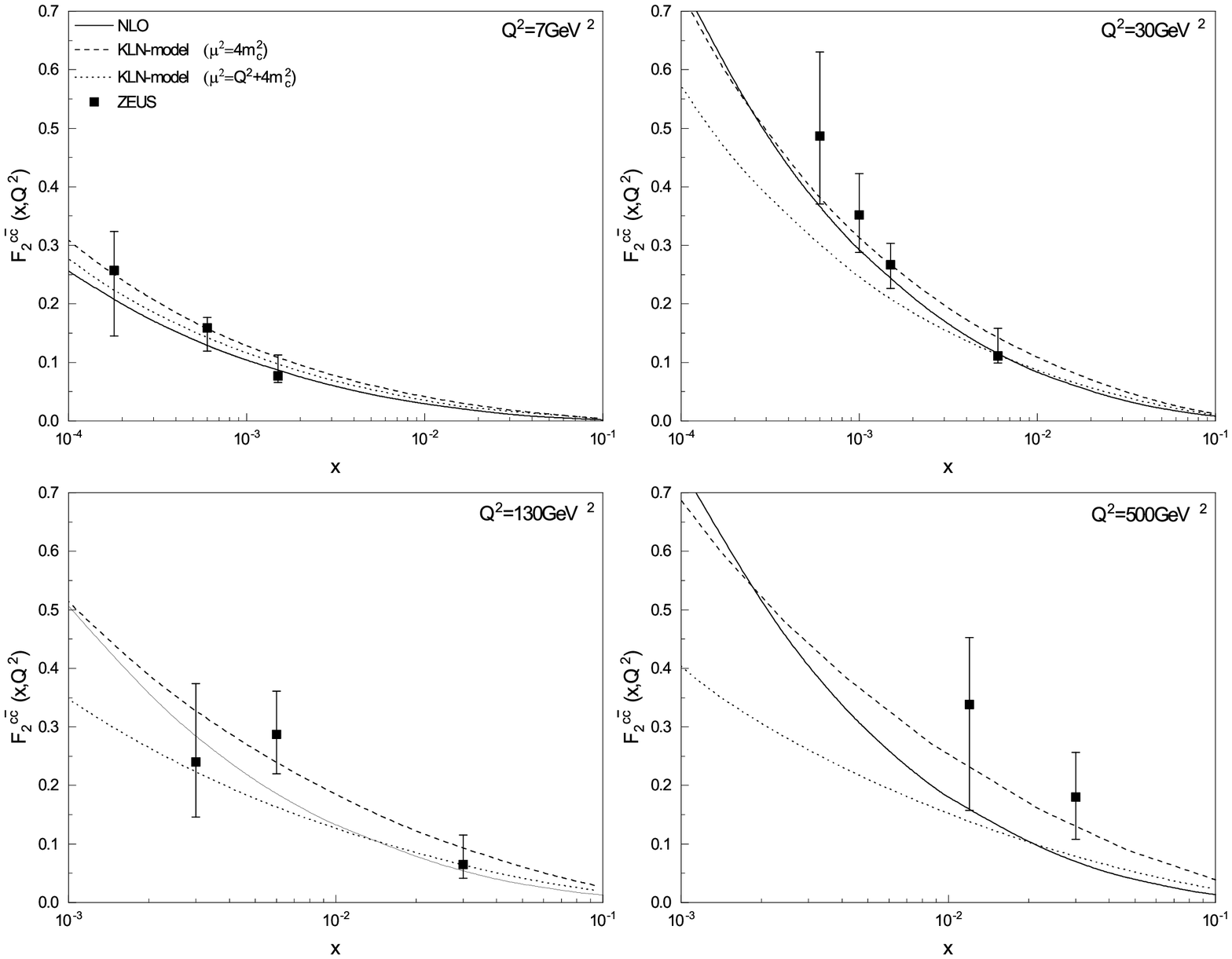}
\end{center}
\begin{center}
\selectfont{\label*{Figure (5): The charm structure function $F^{c\bar{c}}_{2} $ at the NLO approximation at $Q^{2} = 7, 30, 130$ and $500$ $GeV^{2}$. These results have been compared with the ZEUS data \cite{26} and with the results from KLN model \cite{29} obtained for $\mu^{2}=Q^{2}+4m_{c}^{2}$ and $\mu^{2}= 4m_{c}^{2}$.}}
\end{center}
\end{figure}
\begin{figure}[h]
\begin{center}
\includegraphics[width=.65\textwidth]{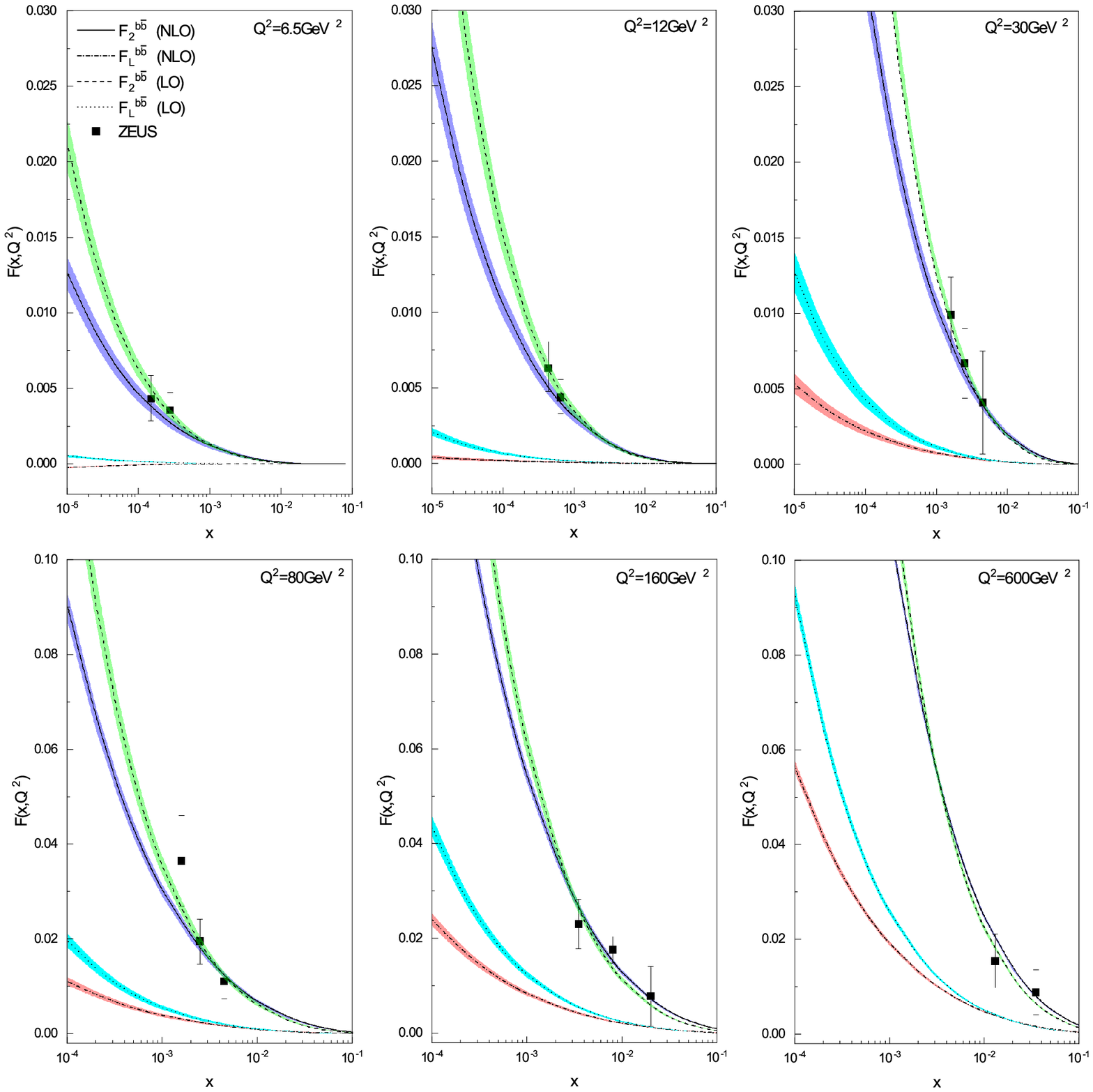}
\end{center}
\begin{center}
\selectfont{\label*{Figure (6): The scale evolution of the beauty quark structure functions $F^{b\bar{b}}_{2} $ and $F^{b\bar{b}}_{L} $. The dashed and solid curves respectively show our numerical results at the LO and NLO approximations which $F^{b\bar{b}}_{2} $ are compared with the ZEUS  \cite{27} data (solid points).}}
\end{center}
\end{figure}

\begin{figure}[h]
\begin{center}
\includegraphics[width=.48\textwidth]{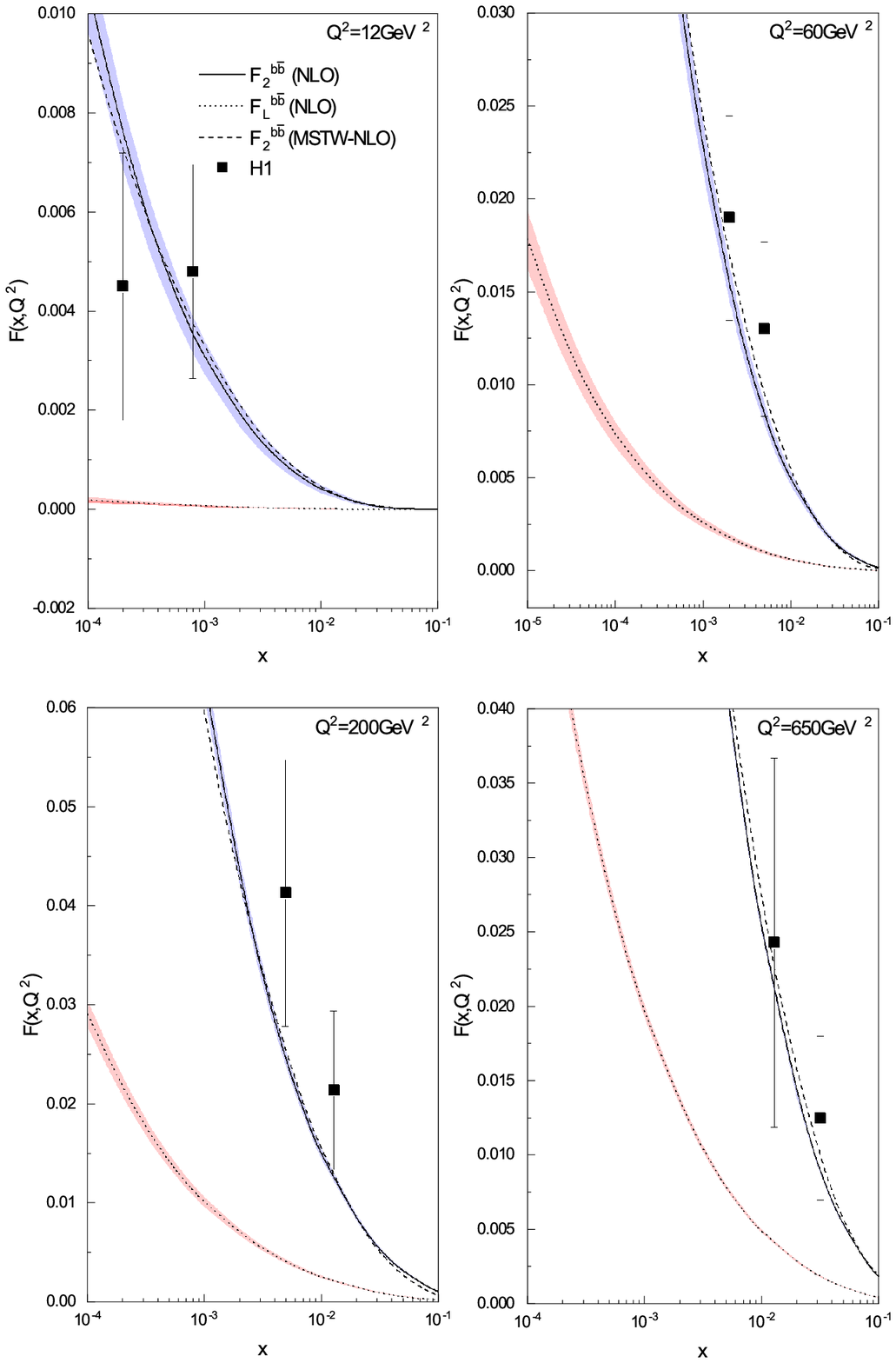}
\end{center}
\begin{center}
\selectfont{\label*{Figure (7): The beauty structure function $F^{b\bar{b}}_{2}$ compared to data from H1 \cite{28} and MSTW2008 at the NLO approximation \cite{32}.}}
\end{center}
\end{figure}
\begin{figure}[h]
\begin{center}
\includegraphics[width=.5\textwidth]{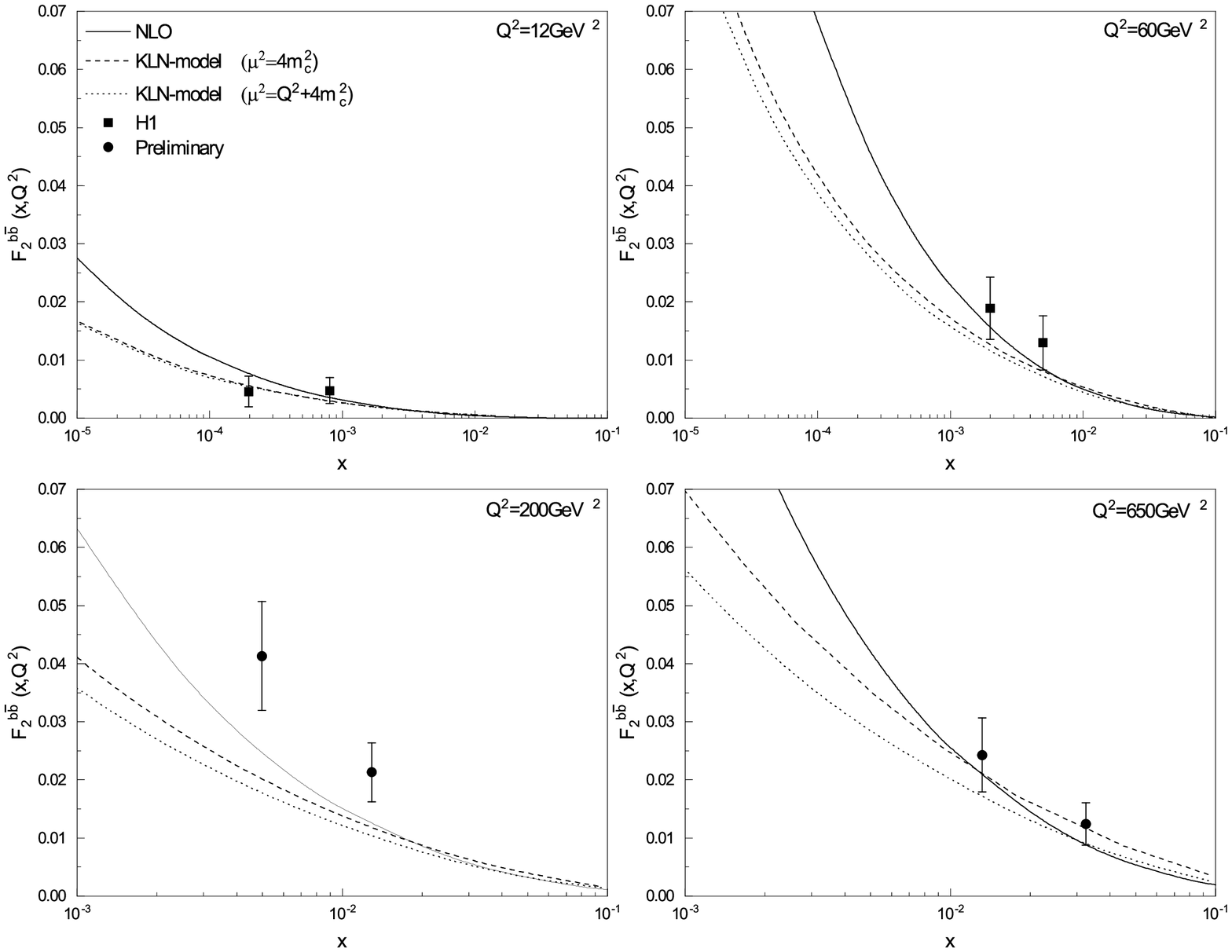}
\end{center}
\begin{center}
\selectfont{\label*{Figure (8): A comparison between the beauty quark structure function $F^{b\bar{b}}_{2} $ at the H1 collaboration \cite{28}, our numerical results at the NLO approximation and the results from KLN model \cite{29}.}}
\end{center}
\end{figure}
\begin{figure}[h]
\begin{center}
\includegraphics[width=.65\textwidth]{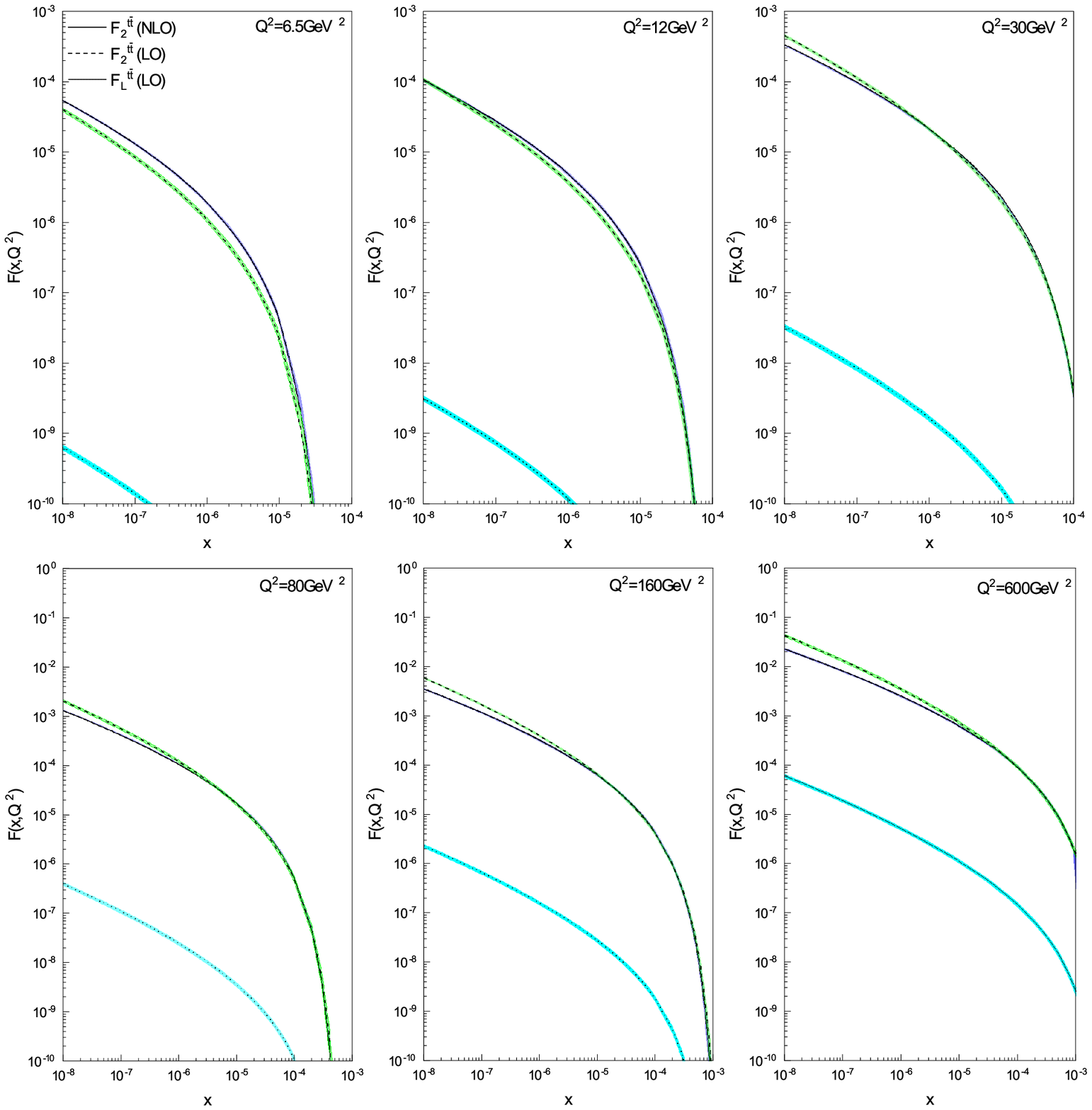}
\end{center}
\begin{center}
\selectfont{\label*{Figure (9): The results of  $F^{t\bar{t}}_{2} $ at $Q^{2}$ values between $6.5$ and $600$ $GeV^{2}$ as a function of $x$ at the LO and NLO approximations and the results of  $F^{t\bar{t}}_{L}$ at the LO approximation. }}
\end{center}
\end{figure}
\begin{figure}[h]
\begin{center}
\includegraphics[width=.40\textwidth]{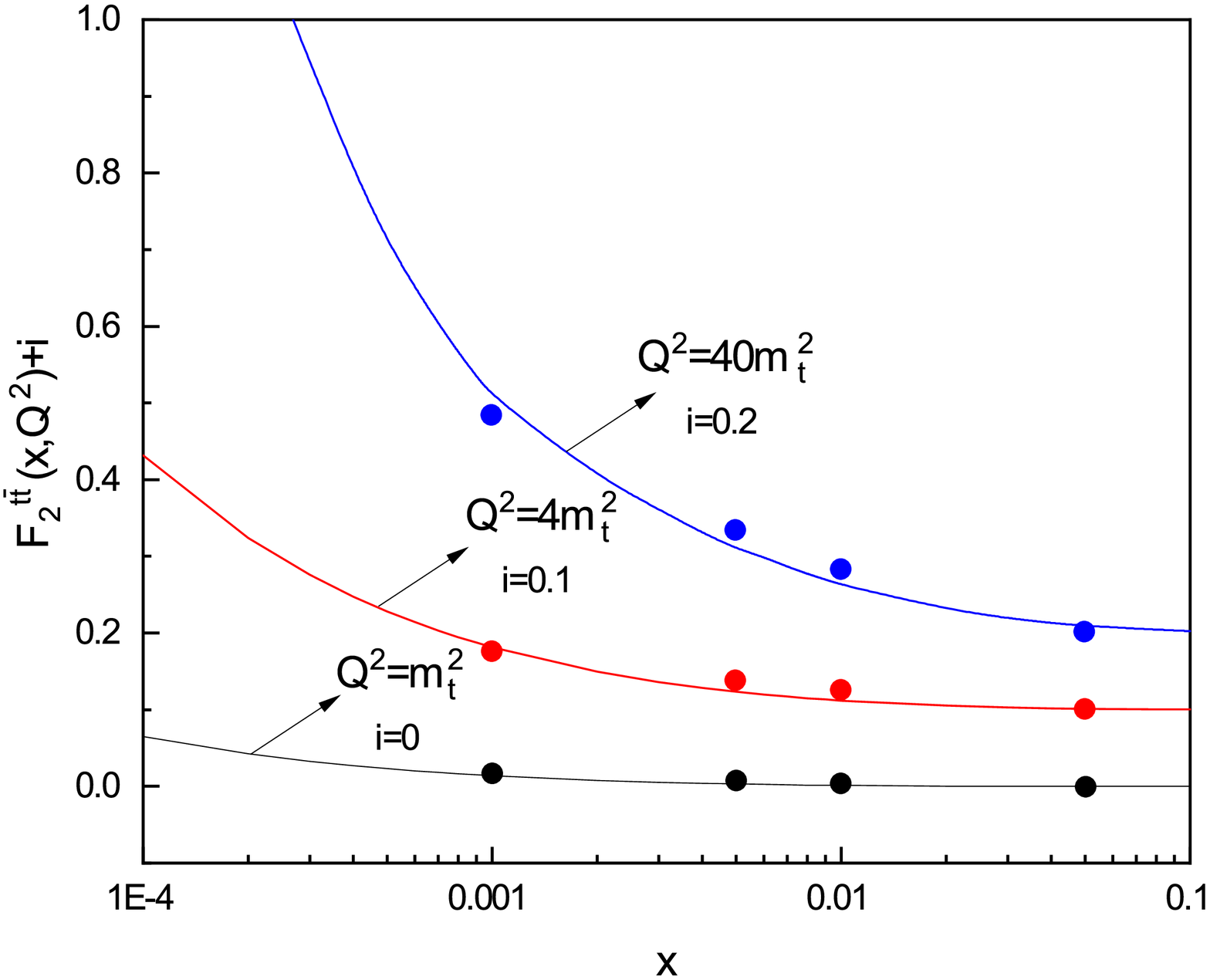}
\end{center}
\begin{center}
\selectfont{\label*{Figure (10): The results of  $F^{t\bar{t}}_{2} $ at $Q^{2}=m_{t}^{2}$, $4m_{t}^{2}$ and $40m_{t}^{2}$ at the NLO approximation compared with the results of Ref. \cite{21}.  }}
\end{center}
\end{figure}
\begin{figure}[h]
\begin{center}
\includegraphics[width=.5\textwidth]{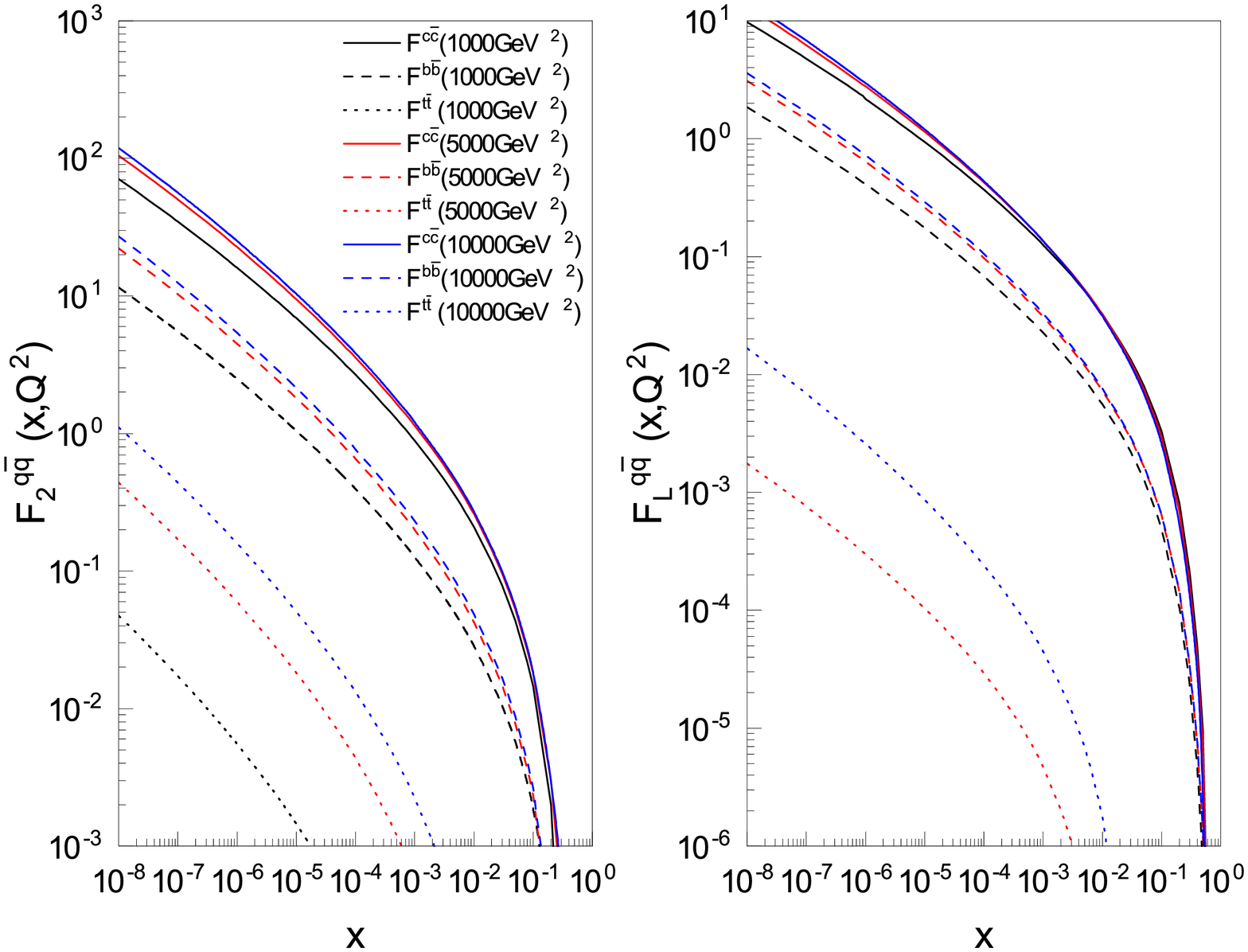}
\end{center}
\begin{center}
\selectfont{\label*{Figure (11): The heavy quark structure functions $F^{q\bar{q}}_{2} $ and $F^{q\bar{q}}_{L}$ (with $q=c$, $b$ and $t$) at the NLO approximation at $Q^{2} = 1000, 5000$ and $10000$ $GeV^{2}$. }}
\end{center}
\end{figure} 

\begin{figure}[h]
\begin{center}
\includegraphics[width=.4\textwidth]{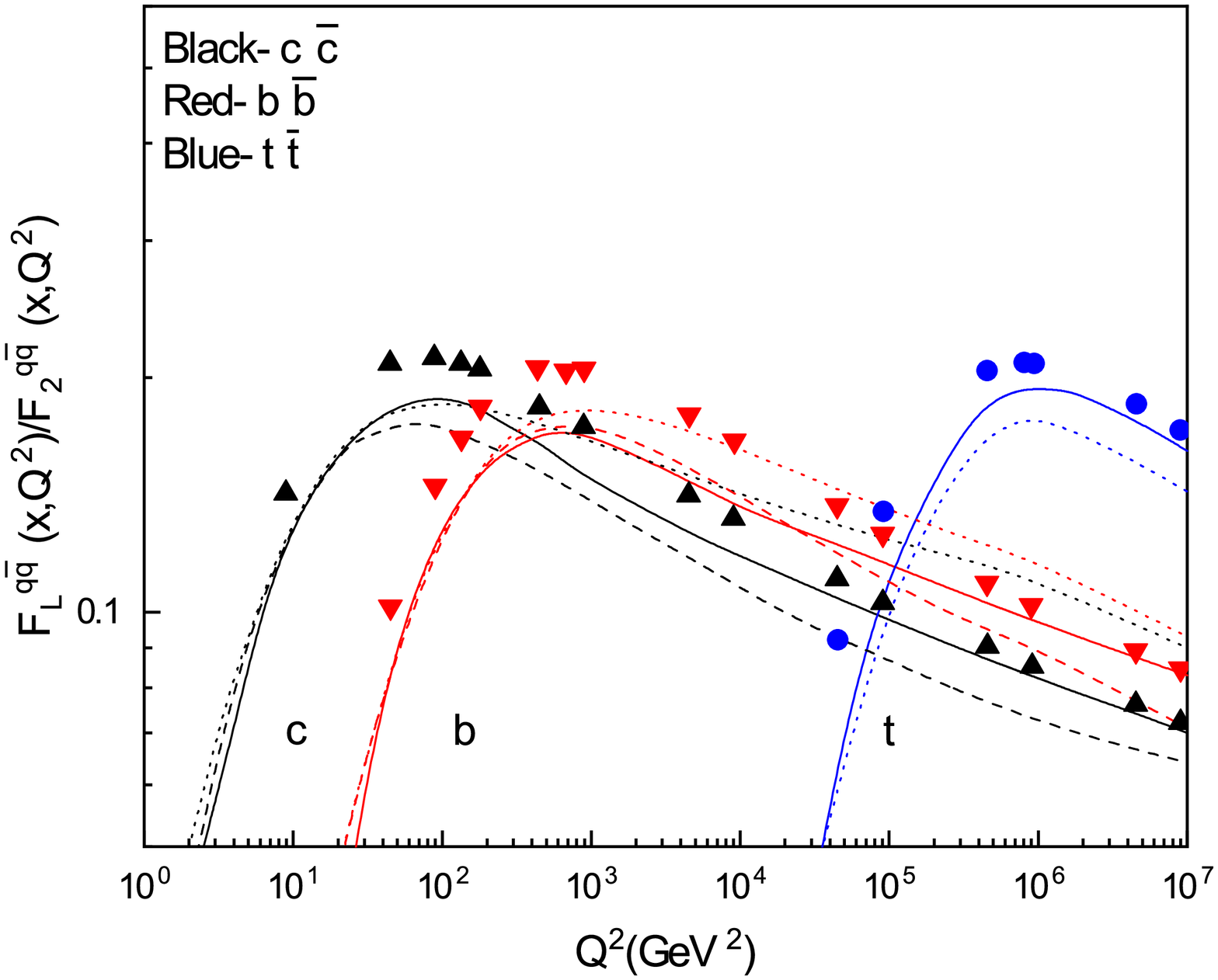}
\end{center}
\begin{center}
\selectfont{\label*{Figure (12): The ratios $R^{c\bar{c}}$, $R^{b\bar{b}}$ and $R^{t\bar{t}}$ evaluated as functions of $Q^{2}$ at the NLO approximation (solid-curves). For comparison, the results of these ratios obtained by Ref. \cite{21} and Ref. \cite{22} are shown by symbols and dashed-lines, respectively. }}
\end{center}
\end{figure}
In conclusion, we have presented the heavy quarks contributions to the proton structure functions $F_{2}^{q\bar{q}}$ and $F_{L}^{q\bar{q}}$ by using Dokshitzer-Gribov-Lipatov-Altarelli-Parisi evolution equations and by utilizing a suitable fit for the  heavy quarks coefficient functions at the NLO approximation. Indeed, there are various methods presented the charm and beauty quarks structure functions, but in this paper, we have shown that the Laplace transform method is the suitable and alternative scheme to solve the DGLAP evolution equations and Eq. (2). Also, the obtained equations are general and require only a knowledge of the parton distribution functions $F_{s}(x)$, $G(x)$ at the starting value $Q^{2}_{0}$. The comparisons have shown that our numerical results of the charm and beauty structure functions, $F_{2}^{c\bar{c}}$ and $F_{2}^{b\bar{b}}$, are in agreement with the H1 and ZEUS data \cite{26,27,28,281} well within errors. Furthermore, in this paper, we compared the functions of $F_{2}^{c\bar{c}}$ and $F_{2}^{b\bar{b}}$ with those obtained from the KLN model \cite{29} and with the MSTW2008 predictions at the NOL approximation  \cite{32}. Moreover,  we compared the results of  $F^{t\bar{t}}_{2} $ at $Q^{2}=m_{t}^{2}$, $4m_{t}^{2}$ and $40m_{t}^{2}$ at the NLO approximation with the results of Ref. \cite{21}. Also,  we have obtained the ratio $R^{q\bar{q}} = F^{q\bar{q}}_{L}/F^{q\bar{q}}_{2}$ for the heavy flavour contributions to the proton structure functions and compared them with the results from Refs. \cite{21,22}. Finally, we note that so far no results have been provided for the top quark longitudinal structure function and only in recent articles the ratio $ F_{L}^{t\bar{t}}/ F_{2}^{t\bar{t}}$ has been obtained and presented, so,  we only compared the numerical result of this ratio with the results of Ref. \cite{21,22}.

\pagebreak
\begin{appendix}
\setcounter{equation}{0}
\section*{ Appendix: THE COEFFICIENT FUNCTIONS}
\begin{widetext}
The expanded coefficients $H^{(1)}_{2,g}$ and $H^{(1)}_{L,g}$ in Eq. (4) at the LO approximation are as follows:
$$
H^{(1)}_{L,g}(z,\xi)=z \bigg((a^2+2. a-3) z^2+(0.25 a^3+0.25 a^2+0.75 a-1.25) z^3+(0.104167 a^4+0.0833333 a^3+0.125 a^2+0.416667 a
$$
$$
-0.729167) z^4+(0.0546875 a^5+0.0390625 a^4+0.046875 a^3+0.078125 a^2+0.273438 a-0.492188) z^5+(0.0328125 a^6+0.021875 a^5
$$
$$
+0.0234375 a^4+0.03125 a^3+0.0546875 a^2+0.196875 a-0.360938) z^6+(0.0214844 a^7+0.0136719 a^6+0.0136719 a^5+0.016276 a^4
$$
$$
+0.0227865 a^3+0.0410156 a^2+0.150391 a-0.279297) z^7+(0.0149623 a^8+0.00920759 a^7+0.00878906 a^6+0.00976563 a^5
$$
$$
+0.012207 a^4+0.0175781 a^3+0.0322266 a^2+0.119699 a-0.224435) z^8+(0.01091 a^9+0.00654602 a^8+0.00604248 a^7+0.00640869
$$
$$
 \times a^6+0.00747681 a^5+0.00961304 a^4+0.0140991 a^3+0.0261841 a^2+0.0981903 a-0.185471) z^9+(0.00824314 a^{10}+0.0048489 a^9
$$
$$ 
 +0.00436401 a^8+0.00447591 a^7+0.00498454 a^6+0.00598145 a^5+0.00783285 a^4+0.0116374 a^3+0.0218201 a^2+0.0824314 a
 $$
 $$
 -0.15662) z^{10}+(-6.77259 a-1.22741) z+(4 a-4) z \ln (0.5 (a-1) z)+(0.00640717 a^{11}+0.00370941 a^{10}+0.00327301 a^9
 $$
 $$
 +0.00327301 a^8+0.00352478 a^7+0.00403748 a^6+0.00493469 a^5+0.00654602 a^4+0.00981903 a^3+0.0185471 a^2+0.0704788 a
$$
$$ 
 -0.13455) z^{11}+(0.00509661 a^{12}+0.00291235 a^{11}+0.00252914 a^{10}+0.00247955 a^9+0.00260353 a^8+0.00288391 a^7+0.00336456 a^6
$$
$$ 
+0.00416565 a^5+0.00557899 a^4+0.00843048 a^3+0.0160179 a^2+0.0611593 a-0.117222) z^{12}+(0.00413283 a^{13}+0.00233595 a^{12}
$$
$$
+0.00200224 a^{11}+0.00193199 a^{10}+0.00198881 a^9+0.00214791 a^8+0.00242329 a^7+0.00286388 a^6+0.00357985 a^5+0.00482996 a^4
$$
$$
+0.00734154 a^3+0.0140157 a^2+0.0537268 a-0.103321) z^{13}+(0.00340618 a^{14}+0.00190746 a^{13}+0.00161719 a^{12}+0.00154018 a^{11}
$$
$$
+0.00156045 a^{10}+0.00165224 a^9+0.00181746 a^8+0.0020771 a^7+0.00247836 a^6+0.0031209 a^5+0.00423551 a^4+0.00646877 a^3
$$
$$
+0.0123985 a^2+0.0476865 a-0.0919668) z^{14}+(0.00284659 a^{15}+0.00158144 a^{14}+0.00132841 a^{13}+0.0012514 a^{12}+0.0012514 a^{11}
$$
$$
+0.00130409 a^{10}+0.00140637 a^9+0.0015671 a^8+0.00180819 a^7+0.00217348 a^6+0.00275308 a^5+0.0037542 a^4+0.00575644 a^3
$$
$$
+0.0110701 a^2+0.0426989 a-0.0825511) z^{15}+(0.0385239 a^{16}+0.0212545 a^{15}+0.0177121 a^{14}+0.0165313 a^{13}+0.0163516 a^{12}
$$
$$
+0.0168188 a^{11}+0.0178515 a^{10}+0.0195017 a^9+0.0219394 a^8+0.0255022 a^7+0.0308345 a^6+0.0392439 a^5+0.0537268 a^4
$$
\begin{equation}
+0.0826565 a^3+0.159409 a^2+0.616382 a-1.19424) z^{16}+8\bigg),
\end{equation}

$$
H^{(1)}_{2,g}(z,\xi)=(0.00963096 a^{17}+0.0335423 a^{16}+0.0212545 a^{15}+0.0191881 a^{14}+0.0190919 a^{13}+0.0199723 a^{12}+0.0216136 a^{11}
$$
$$
+0.0240396 a^{10}+0.0274242 a^9+0.0321122 a^8+0.038717 a^7+0.0483541 a^6+0.0631873 a^5+0.0878226 a^4+0.132841 a^3+0.193948 a^2
$$
$$
+1.62763 a-2.68704) z^{17}+(0.00137585 a^{16}+0.000853977 a^{15}+0.000790719 a^{14}+0.000811805 a^{13}+0.00087598 a^{12}+0.000976092
$$
$$
\times a^{11}+0.00111572 a^{10}+0.00130592 a^9+0.0015671 a^8+0.00193544 a^7+0.00247777 a^6+0.00332872 a^5+0.00479703 a^4
$$
$$
+0.00774905 a^3+0.0156562 a^2+0.0632892 a-0.127954) z^{16}+(-0.0076566 a^{15}-0.00339401 a^{14}-0.00212886 a^{13}-0.0013252 a^{12}
$$
$$
-0.000644952 a^{11}+0.000036782 a^{10}+0.000804156 a^9+0.00174792 a^8+0.00299974 a^7+0.00478724 a^6+0.00756038 a^5+0.0123504 
$$
$$
\times a^4+0.0220663 a^3+0.0484529 a^2+0.2101 a-0.451068) z^{15}+(-0.00829242 a^{14}-0.00359919 a^{13}-0.00216588 a^{12}-0.00121931 a^{11}
$$
$$
-0.000380824 a^{10}+0.000501573 a^9+0.00154701 a^8+0.00290547 a^7+0.00482395 a^6+0.00778367 a^5+0.012883 a^4+0.0232183 a^3
$$
$$
+0.0512913 a^2+0.223388 a-0.481183) z^{14}+(-0.00900263 a^{13}-0.00379795 a^{12}-0.00214926 a^{11}-0.00100652 a^{10}+0.0000625849 a^9
$$
$$
+0.0012542 a^8+0.00275373 a^7+0.00483656 a^6+0.00802338 a^5+0.0134935 a^4+0.0245659 a^3+0.0546415 a^2+0.23914 a-0.516965)
$$
$$
\times z^{13}+(-0.00978974 a^{12}-0.00396807 a^{11}-0.00203385 a^{10}-0.000609557 a^9+0.000813603 a^8+0.00251141 a^7+0.00480852 a^6
$$
$$
+0.00827923 a^5+0.0142032 a^4+0.0261696 a^3+0.0586656 a^2+0.258143 a-0.560224) z^{12}+(-0.0106462 a^{11}-0.00406537 a^{10}
$$
$$
-0.00173569 a^9+0.000115712 a^8+0.00211843 a^7+0.00471039 a^6+0.00854848 a^5+0.0150426 a^4+0.0281181 a^3+0.0636033 a^2
$$
$$
+0.281559 a-0.613632) z^{11}+(-0.0115404 a^{10}-0.00400035 a^9-0.001091 a^8+0.00145467 a^7+0.00448608 a^6+0.00882263 a^5
$$
$$
+0.0160573 a^4+0.0305481 a^3+0.0698242 a^2+0.311178 a-0.681295) z^{10}+(-0.012382 a^9-0.00358473 a^8+0.000239781 a^7
$$
$$
+0.00401815 a^6+0.00907898 a^5+0.0173187 a^4+0.0336812 a^3+0.0779288 a^2+0.3499 a-0.76985) z^9+(-0.0129362 a^8-0.00239781 
$$
$$
\times a^7+0.00302124 a^6+0.009257 a^5+0.0189463 a^4+0.0379028 a^3+0.0889587 a^2+0.402736 a-0.890726) z^8+(-0.0125837 a^7
$$
$$
+0.000585938 a^6+0.00917969 a^5+0.0211589 a^4+0.0439453 a^3+0.104883 a^2+0.479102 a-1.06532) z^7+(-0.00957031 a^6
$$
$$
+0.00820313 a^5+0.0244141 a^4+0.0533854 a^3+0.129883 a^2+0.598828 a-1.33848) z^6+(0.00182292 a^5+0.0299479 a^4+0.0703125 
$$
$$
\times a^3+0.174479 a^2+0.811198 a-1.82109) z^5+(0.0455729 a^4+0.109375 a^3+0.273438 a^2+1.27604 a-2.87109) z^4+(0.291667 a^3
$$
$$
+0.625 a^2+2.875 a-6.45833) z^3+(-0.818147 a^2-8.52259 a-0.886675) z^2+(11.8411 -0.613706 a) z+((a^2+4 a-9) z^2
$$
\begin{equation}
+(6 -2 a) z-2.) \ln (0.5 (a-1) z)-0.613706,
\end{equation}
where $a=1+\frac{4}{\xi}$. The coefficients $h^{(1)}_{2,g}$ and $h^{(1)}_{L,g}$ in Eq. (16) at the LO approximation are as follows:
$$
h^{(1)}_{L,g}(s,a)=\frac{4 \ln \left(\frac{a-1}{a}\right) s}{a^2 (s+3)^2}-\frac{2.77259 s}{a^2 (s+3)^2}-\frac{4 \ln \left(\frac{a-1}{a}\right) s}{a^3 (s+3)^2}+\frac{2.77259 s}{a^3 (s+3)^2}+\frac{12 \ln \left(\frac{a-1}{a}\right)}{a^2 (s+3)^2}+\frac{8}{a^2 (s+2)}-\frac{6.77259}{a^2 (s+3)}-\frac{1.22741}{a^3 (s+3)}
$$
$$
-\frac{12.3178}{a^2 (s+3)^2}-\frac{12 \ln \left(\frac{a-1}{a}\right)}{a^3 (s+3)^2}+\frac{12.3178}{a^3 (s+3)^2}+\frac{1}{a^2 (s+4)}+\frac{2}{a^3 (s+4)}-\frac{3}{a^4 (s+4)}+\frac{0.25}{a^2 (s+5)}+\frac{0.25}{a^3 (s+5)}+\frac{0.75}{a^4 (s+5)}-\frac{1.25}{a^5 (s+5)}
$$
$$
+\frac{0.104167}{a^2 (s+6)}+\frac{0.0833333}{a^3 (s+6)}+\frac{0.125}{a^4 (s+6)}+\frac{0.416667}{a^5 (s+6)}-\frac{0.729167}{a^6 (s+6)}+\frac{0.0546875}{a^2 (s+7)}+\frac{0.0390625}{a^3 (s+7)}+\frac{0.046875}{a^4 (s+7)}+\frac{0.078125}{a^5 (s+7)}+\frac{0.273438}{a^6 (s+7)}
$$
$$
-\frac{0.492188}{a^7 (s+7)}+\frac{0.0328125}{a^2 (s+8)}+\frac{0.021875}{a^3 (s+8)}+\frac{0.0234375}{a^4 (s+8)}+\frac{0.03125}{a^5 (s+8)}+\frac{0.0546875}{a^6 (s+8)}+\frac{0.196875}{a^7 (s+8)}-\frac{0.360938}{a^8 (s+8)}+\frac{0.0214844}{a^2 (s+9)}+\frac{0.0136719}{a^3 (s+9)}
$$
$$
+\frac{0.0136719}{a^4 (s+9)}+\frac{0.016276}{a^5 (s+9)}+\frac{0.0227865}{a^6 (s+9)}+\frac{0.0410156}{a^7 (s+9)}+\frac{0.150391}{a^8 (s+9)}-\frac{0.279297}{a^9 (s+9)}+\frac{0.0149623}{a^2 (s+10)}+\frac{0.00920759}{a^3 (s+10)}+\frac{0.00878906}{a^4 (s+10)}+\frac{0.00976563}{a^5 (s+10)}
$$
$$
+\frac{0.012207}{a^6 (s+10)}+\frac{0.0175781}{a^7 (s+10)}+\frac{0.0322266}{a^8 (s+10)}+\frac{0.119699}{a^9 (s+10)}-\frac{0.224435}{a^{10} (s+10)}+\frac{0.01091}{a^2 (s+11)}+\frac{0.00654602}{a^3 (s+11)}+\frac{0.00604248}{a^4 (s+11)}+\frac{0.00640869}{a^5 (s+11)}
$$
$$
+\frac{0.00747681}{a^6 (s+11)}+\frac{0.00961304}{a^7 (s+11)}+\frac{0.0140991}{a^8 (s+11)}+\frac{0.0261841}{a^9 (s+11)}+\frac{0.0981903}{a^{10} (s+11)}-\frac{0.185471}{a^{11} (s+11)}+\frac{0.00824314}{a^2 (s+12)}+\frac{0.0048489}{a^3 (s+12)}+\frac{0.00436401}{a^4 (s+12)}
$$
$$
+\frac{0.00447591}{a^5 (s+12)}+\frac{0.00498454}{a^6 (s+12)}+\frac{0.00598145}{a^7 (s+12)}+\frac{0.00783285}{a^8 (s+12)}+\frac{0.0116374}{a^9 (s+12)}+\frac{0.0218201}{a^{10} (s+12)}+\frac{0.0824314}{a^{11} (s+12)}-\frac{0.15662}{a^{12} (s+12)}+\frac{0.00640717}{a^2 (s+13)}
$$
$$
+\frac{0.00370941}{a^3 (s+13)}+\frac{0.00327301}{a^4 (s+13)}+\frac{0.00327301}{a^5 (s+13)}+\frac{0.00352478}{a^6 (s+13)}+\frac{0.00403748}{a^7 (s+13)}+\frac{0.00493469}{a^8 (s+13)}+\frac{0.00654602}{a^9 (s+13)}+\frac{0.00981903}{a^{10} (s+13)}+\frac{0.0185471}{a^{11} (s+13)}
$$
$$
+\frac{0.0704788}{a^{12} (s+13)}-\frac{0.13455}{a^{13} (s+13)}+\frac{0.00509661}{a^2 (s+14)}+\frac{0.00291235}{a^3 (s+14)}+\frac{0.00252914}{a^4 (s+14)}+\frac{0.00247955}{a^5 (s+14)}+\frac{0.00260353}{a^6 (s+14)}+\frac{0.00288391}{a^7 (s+14)}+\frac{0.00336456}{a^8 (s+14)}
$$
$$
+\frac{0.00416565}{a^9 (s+14)}+\frac{0.00557899}{a^{10} (s+14)}+\frac{0.00843048}{a^{11} (s+14)}+\frac{0.0160179}{a^{12} (s+14)}+\frac{0.0611593}{a^{13} (s+14)}-\frac{0.117222}{a^{14} (s+14)}+\frac{0.00413283}{a^2 (s+15)}+\frac{0.00233595}{a^3 (s+15)}+\frac{0.00200224}{a^4 (s+15)}
$$
$$
+\frac{0.00193199}{a^5 (s+15)}+\frac{0.00198881}{a^6 (s+15)}+\frac{0.00214791}{a^7 (s+15)}+\frac{0.00242329}{a^8 (s+15)}+\frac{0.00286388}{a^9 (s+15)}+\frac{0.00357985}{a^{10} (s+15.)}+\frac{0.00482996}{a^{11} (s+15.)}+\frac{0.00734154}{a^{12} (s+15)}
$$
$$
+\frac{0.0140157}{a^{13} (s+15)}+\frac{0.0537268}{a^{14} (s+15)}-\frac{0.103321}{a^{15} (s+15)}+\frac{0.00340618}{a^2 (s+16)}+\frac{0.00190746}{a^3 (s+16)}+\frac{0.00161719}{a^4 (s+16)}+\frac{0.00154018}{a^5 (s+16)}+\frac{0.00156045}{a^6 (s+16)}+\frac{0.00165224}{a^7 (s+16)}
$$
$$
+\frac{0.00181746}{a^8 (s+16)}+\frac{0.0020771}{a^9 (s+16)}+\frac{0.00247836}{a^{10} (s+16)}+\frac{0.0031209}{a^{11} (s+16)}+\frac{0.00423551}{a^{12} (s+16)}+\frac{0.00646877}{a^{13} (s+16)}+\frac{0.0123985}{a^{14} (s+16)}+\frac{0.0476865}{a^{15} (s+16)}
$$
$$
-\frac{0.0919668}{a^{16} (s+16)}+\frac{0.00284659}{a^2 (s+17)}+\frac{0.00158144}{a^3 (s+17)}+\frac{0.00132841}{a^4 (s+17)}+\frac{0.0012514}{a^5 (s+17)}+\frac{0.0012514}{a^6 (s+17)}+\frac{0.00130409}{a^7 (s+17)}+\frac{0.00140637}{a^8 (s+17)}+\frac{0.0015671}{a^9 (s+17)}
$$
$$
+\frac{0.00180819}{a^{10} (s+17)}+\frac{0.00217348}{a^{11} (s+17)}+\frac{0.00275308}{a^{12} (s+17)}+\frac{0.0037542}{a^{13} (s+17)}+\frac{0.00575644}{a^{14} (s+17)}+\frac{0.0110701}{a^{15} (s+17)}+\frac{0.0426989}{a^{16} (s+17)}-\frac{0.0825511}{a^{17} (s+17)}+\frac{0.0385239}{a^2 (s+18)}
$$
$$
+\frac{0.0212545}{a^3 (s+18)}+\frac{0.0177121}{a^4 (s+18)}+\frac{0.0165313}{a^5 (s+18)}+\frac{0.0163516}{a^6 (s+18)}+\frac{0.0168188}{a^7 (s+18)}+\frac{0.0178515}{a^8 (s+18)}+\frac{0.0195017}{a^9 (s+18)}+\frac{0.0219394}{a^{10} (s+18)}+\frac{0.0255022}{a^{11} (s+18)}
$$
\begin{equation}
+\frac{0.0308345}{a^{12} (s+18)}+\frac{0.0392439}{a^{13} (s+18)}+\frac{0.0537268}{a^{14} (s+18)}+\frac{0.0826565}{a^{15} (s+18.)}+\frac{0.159409}{a^{16} (s+18)}+\frac{0.616382}{a^{17} (s+18)}-\frac{1.19424}{a^{18} (s+18)},
\end{equation}

$$
h^{(1)}_{2,g}(s,a)=\frac{6 \ln \left(\frac{a-1}{a}\right) s}{a^2 (s+2)^2}+\frac{\ln \left(\frac{a-1}{a}\right) s}{a (s+3)^2}+\frac{4 \ln \left(\frac{a-1}{a}\right) s}{a^2 (s+3)^2}-\frac{2 \ln \left(\frac{a-1}{a}\right) s}{a (s+1)^2}+\frac{1.38629 s}{a (s+1)^2}-\frac{2 \ln \left(\frac{a-1}{a}\right) s}{a (s+2)^2}+\frac{1.38629 s}{a (s+2)^2}-\frac{4.15888 s}{a^2 (s+2)^2}
$$
$$
-\frac{0.693147 s}{a (s+3)^2}-\frac{2.77259 s}{a^2 (s+3)^2}-\frac{9 \ln \left(\frac{a-1}{a}\right) s}{a^3 (s+3)^2}+\frac{6.23832 s}{a^3 (s+3)^2}+\frac{12 \ln \left(\frac{a-1}{a}\right)}{a^2 (s+2)^2}+\frac{3 \ln \left(\frac{a-1}{a}\right)}{a (s+3)^2}+\frac{12 \ln \left(\frac{a-1}{a}\right)}{a^2 (s+3)^2}-\frac{0.613706}{a (s+1)}-\frac{2 \ln \left(\frac{a-1}{a}\right)}{a (s+1)^2}
$$
$$
+\frac{3.38629}{a (s+1)^2}-\frac{0.613706}{a (s+2)}+\frac{11.8411}{a^2 (s+2)}-\frac{4. \ln \left(\frac{a-1}{a}\right)}{a (s+2)^2}+\frac{4.77259}{a (s+2)^2}-\frac{14.3178}{a^2 (s+2)^2}-\frac{0.818147}{a (s+3)}-\frac{8.52259}{a^2 (s+3)}-\frac{0.886675}{a^3 (s+3)}-\frac{3.07944}{a (s+3)^2}
$$
$$
-\frac{12.3178}{a^2 (s+3)^2}-\frac{27 \ln \left(\frac{a-1}{a}\right)}{a^3 (s+3)^2}+\frac{27.715}{a^3 (s+3)^2}+\frac{0.291667}{a (s+4)}+\frac{0.625}{a^2 (s+4)}+\frac{2.875}{a^3 (s+4)}-\frac{6.45833}{a^4 (s+4)}+\frac{0.0455729}{a (s+5)}+\frac{0.109375}{a^2 (s+5)}+\frac{0.273438}{a^3 (s+5)}
$$
$$
+\frac{1.27604}{a^4 (s+5)}-\frac{2.87109}{a^5 (s+5)}+\frac{0.00182292}{a (s+6)}+\frac{0.0299479}{a^2 (s+6)}+\frac{0.0703125}{a^3 (s+6)}+\frac{0.174479}{a^4 (s+6)}+\frac{0.811198}{a^5 (s+6)}-\frac{1.82109}{a^6 (s+6)}-\frac{0.00957031}{a (s+7)}+\frac{0.00820313}{a^2 (s+7)}
$$
$$
+\frac{0.0244141}{a^3 (s+7)}+\frac{0.0533854}{a^4 (s+7)}+\frac{0.129883}{a^5 (s+7)}+\frac{0.598828}{a^6 (s+7)}-\frac{1.33848}{a^7 (s+7)}-\frac{0.0125837}{a (s+8)}+\frac{0.000585938}{a^2 (s+8)}+\frac{0.00917969}{a^3 (s+8)}+\frac{0.0211589}{a^4 (s+8)}+\frac{0.0439453}{a^5 (s+8)}
$$
$$
+\frac{0.104883}{a^6 (s+8)}+\frac{0.479102}{a^7 (s+8)}-\frac{1.06532}{a^8 (s+8)}-\frac{0.0129362}{a (s+9)}-\frac{0.00239781}{a^2 (s+9)}+\frac{0.00302124}{a^3 (s+9)}+\frac{0.009257}{a^4 (s+9)}+\frac{0.0189463}{a^5 (s+9)}+\frac{0.0379028}{a^6 (s+9)}+\frac{0.0889587}{a^7 (s+9)}
$$
$$
+\frac{0.402736}{a^8 (s+9)}-\frac{0.890726}{a^9 (s+9)}-\frac{0.012382}{a (s+10)}-\frac{0.00358473}{a^2 (s+10)}+\frac{0.000239781}{a^3 (s+10)}+\frac{0.00401815}{a^4 (s+10)}+\frac{0.00907898}{a^5 (s+10)}+\frac{0.0173187}{a^6 (s+10)}+\frac{0.0336812}{a^7 (s+10)}
$$
$$
+\frac{0.0779288}{a^8 (s+10)}+\frac{0.3499}{a^9 (s+10)}-\frac{0.76985}{a^{10} (s+10)}-\frac{0.0115404}{a (s+11)}-\frac{0.00400035}{a^2 (s+11)}-\frac{0.001091}{a^3 (s+11)}+\frac{0.00145467}{a^4 (s+11)}+\frac{0.00448608}{a^5 (s+11)}+\frac{0.00882263}{a^6 (s+11)}
$$
$$
+\frac{0.0160573}{a^7 (s+11)}+\frac{0.0305481}{a^8 (s+11)}+\frac{0.0698242}{a^9 (s+11)}+\frac{0.311178}{a^{10} (s+11)}-\frac{0.681295}{a^{11} (s+11)}-\frac{0.0106462}{a (s+12)}-\frac{0.00406537}{a^2 (s+12)}-\frac{0.00173569}{a^3 (s+12)}+\frac{0.000115712}{a^4 (s+12)}
$$
$$
+\frac{0.00211843}{a^5 (s+12)}+\frac{0.00471039}{a^6 (s+12)}+\frac{0.00854848}{a^7 (s+12)}+\frac{0.0150426}{a^8 (s+12)}+\frac{0.0281181}{a^9 (s+12)}+\frac{0.0636033}{a^{10} (s+12)}+\frac{0.281559}{a^{11} (s+12)}-\frac{0.613632}{a^{12} (s+12)}-\frac{0.00978974}{a (s+13)}
$$
$$
-\frac{0.00396807}{a^2 (s+13)}-\frac{0.00203385}{a^3 (s+13)}-\frac{0.000609557}{a^4 (s+13)}+\frac{0.000813603}{a^5 (s+13)}+\frac{0.00251141}{a^6 (s+13)}+\frac{0.00480852}{a^7 (s+13)}+\frac{0.00827923}{a^8 (s+13)}+\frac{0.0142032}{a^9 (s+13)}+\frac{0.0261696}{a^{10} (s+13)}
$$
$$
+\frac{0.0586656}{a^{11} (s+13)}+\frac{0.258143}{a^{12} (s+13)}-\frac{0.560224}{a^{13} (s+13)}-\frac{0.00900263}{a (s+14)}-\frac{0.00379795}{a^2 (s+14)}-\frac{0.00214926}{a^3 (s+14)}-\frac{0.00100652}{a^4 (s+14)}+\frac{0.0000625849}{a^5 (s+14)}+\frac{0.0012542}{a^6 (s+14)}
$$
$$
+\frac{0.00275373}{a^7 (s+14)}+\frac{0.00483656}{a^8 (s+14)}+\frac{0.00802338}{a^9 (s+14)}+\frac{0.0134935}{a^{10} (s+14)}+\frac{0.0245659}{a^{11} (s+14)}+\frac{0.0546415}{a^{12} (s+14)}+\frac{0.23914}{a^{13} (s+14)}-\frac{0.516965}{a^{14} (s+14)}
$$
$$
-\frac{0.00829242}{a (s+15)}-\frac{0.00359919}{a^2 (s+15)}-\frac{0.00216588}{a^3 (s+15)}-\frac{0.00121931}{a^4 (s+15)}-\frac{0.000380824}{a^5 (s+15)}+\frac{0.000501573}{a^6 (s+15)}+\frac{0.00154701}{a^7 (s+15)}+\frac{0.00290547}{a^8 (s+15)}+\frac{0.00482395}{a^9 (s+15)}
$$
$$
+\frac{0.00778367}{a^{10} (s+15)}+\frac{0.012883}{a^{11} (s+15)}+\frac{0.0232183}{a^{12} (s+15)}+\frac{0.0512913}{a^{13} (s+15)}+\frac{0.223388}{a^{14} (s+15)}-\frac{0.481183}{a^{15} (s+15)}-\frac{0.0076566}{a (s+16)}-\frac{0.00339401}{a^2 (s+16)}-\frac{0.00212886}{a^3 (s+16)}
$$
$$
-\frac{0.0013252}{a^4 (s+16)}-\frac{0.000644952}{a^5 (s+16)}+\frac{0.000036782}{a^6 (s+16)}+\frac{0.000804156}{a^7 (s+16)}+\frac{0.00174792}{a^8 (s+16)}+\frac{0.00299974}{a^9 (s+16)}+\frac{0.00478724}{a^{10} (s+16)}+\frac{0.00756038}{a^{11} (s+16)}+\frac{0.0123504}{a^{12} (s+16)}
$$
$$
+\frac{0.0220663}{a^{13} (s+16)}+\frac{0.0484529}{a^{14} (s+16)}+\frac{0.2101}{a^{15} (s+16)}-\frac{0.451068}{a^{16} (s+16)}+\frac{0.00137585}{a (s+17)}+\frac{0.000853977}{a^2 (s+17)}+\frac{0.000790719}{a^3 (s+17)}+\frac{0.000811805}{a^4 (s+17)}+\frac{0.00087598}{a^5 (s+17)}
$$
$$
+\frac{0.000976092}{a^6 (s+17)}+\frac{0.00111572}{a^7 (s+17)}+\frac{0.00130592}{a^8 (s+17)}+\frac{0.0015671}{a^9 (s+17)}+\frac{0.00193544}{a^{10} (s+17)}+\frac{0.00247777}{a^{11} (s+17)}+\frac{0.00332872}{a^{12} (s+17)}+\frac{0.00479703}{a^{13} (s+17)}+\frac{0.00774905}{a^{14} (s+17)}
$$
$$
+\frac{0.0156562}{a^{15} (s+17)}+\frac{0.0632892}{a^{16} (s+17)}-\frac{0.127954}{a^{17} (s+17)}+\frac{0.00963096}{a (s+18)}+\frac{0.0335423}{a^2 (s+18)}+\frac{0.0212545}{a^3 (s+18)}+\frac{0.0191881}{a^4 (s+18)}+\frac{0.0190919}{a^5 (s+18)}+\frac{0.0199723}{a^6 (s+18)}
$$
$$
+\frac{0.0216136}{a^7 (s+18)}+\frac{0.0240396}{a^8 (s+18)}+\frac{0.0274242}{a^9 (s+18)}+\frac{0.0321122}{a^{10} (s+18)}+\frac{0.038717}{a^{11} (s+18)}+\frac{0.0483541}{a^{12} (s+18)}+\frac{0.0631873}{a^{13} (s+18)}+\frac{0.0878226}{a^{14} (s+18)}+\frac{0.132841}{a^{15} (s+18)}
$$
\begin{equation}
+\frac{0.193948}{a^{16} (s+18)}+\frac{1.62763}{a^{17} (s+18)}-\frac{2.68704}{a^{18} (s+18)}.
\end{equation}

\end{widetext}

\end{appendix}


\end{document}